\documentclass[sigconf, nonacm]{acmart}

\usepackage{amsmath}
\usepackage[prologue,dvipsnames,svgnames]{xcolor}
\usepackage[linesnumbered,vlined,boxed,ruled]{algorithm2e}
\usepackage{caption}
\usepackage{enumitem}
\usepackage{lipsum}
\usepackage{outlines}
\usepackage{subcaption}
\usepackage{xspace}
\usepackage{pifont}
\usepackage{graphicx}
\usepackage{tikz}
\usepackage{multirow}
\usepackage{balance}

\makeatletter
\@namedef{ver@lineno.sty}{9999/12/31}
\@namedef{opt@lineno.sty}{}
\makeatother
\usepackage[frozencache,cachedir=.]{minted}

\usepackage{colorprofiles}
\usepackage[a-2b,mathxmp]{pdfx}[2018/12/22]

\newcommand*\circled[1]{\tikz[baseline=(char.base)]{
            \node[shape=circle,draw,inner sep=1pt] (char) {#1};}}

\newenvironment{vinlist}
{\begin{itemize}[leftmargin=1em]
  \setlength{\itemsep}{0pt}
  \setlength{\parsep}{0pt}
  \setlength{\parskip}{0pt}
  \setlength{\topsep}{0pt}
  \setlength{\partopsep}{0pt}
  }
{\end{itemize}}

\newcommand\vldbdoi{10.14778/3750601.3750615}
\newcommand\vldbpages{4910 - 4923}
\newcommand\vldbvolume{18}
\newcommand\vldbissue{12}
\newcommand\vldbyear{2025}
\newcommand\vldbauthors{\authors}
\newcommand\vldbtitle{\shorttitle} 
\newcommand\vldbavailabilityurl{https://github.com/microsoft/FASTER/tree/main/cc}
\newcommand\vldbpagestyle{empty}

\newcommand{\faster}{FASTER\xspace}
\newcommand{\hlog}{HybridLog\xspace}
\newcommand{\sys}{F2\xspace}

\newcommand{\bsubsection}[1]{\vspace{.5ex}\noindent\textbf{#1}}
\newcommand{\fbsubsection}[1]{\noindent\textbf{#1}}

\makeatletter
\def\blfootnote{\xdef\@thefnmark{}\@footnotetext}
\makeatother

\begin{document}

\title{From FASTER to F2: Evolving Concurrent Key-Value Store Designs for Large Skewed Workloads}

\author{Konstantinos Kanellis}
\authornote{Work started during internship at Microsoft Research.\vspace{1ex}}
\affiliation{
  \institution{University of Wisconsin-Madison}
  \city{Madison}
  \state{WI}
}
\email{kkanellis@cs.wisc.edu}

\author{Badrish Chandramouli}
\affiliation{
  \institution{Microsoft Research}
  \city{Redmond}
  \state{WA}
}
\email{badrishc@microsoft.com}

\author{Ted Hart}
\affiliation{
  \institution{Microsoft Research}
  \city{Redmond}
  \state{WA}
}
\email{tedhar@microsoft.com}

\author{Shivaram Venkataraman}
\affiliation{
  \institution{University of Wisconsin-Madison}
  \city{Madison}
  \state{WI}
}
\email{shivaram@cs.wisc.edu}

\begin{abstract}
Modern large-scale services such as search engines, messaging platforms, and serverless functions, rely on key-value (KV) stores to maintain high performance at scale.
When such services are deployed in constrained memory environments, they present challenging requirements: point operations requiring high throughput, working sets \textit{much larger} than main memory, and natural \textit{skew} in key access patterns.
Traditional KV stores, based on LSM- and B-Trees, have been widely used to handle such use cases, but they often suffer from suboptimal use of modern hardware resources. The FASTER project, developed as a high-performance open-source KV storage library, has demonstrated remarkable success in both in-memory and hybrid storage environments. However, when tasked with serving large skewed workloads, it faced challenges, including high indexing and compactions overheads, and inefficient management of non-overlapping read-hot and write-hot working sets.

In this paper, we introduce \sys (for FASTER v2), an evolution of \faster designed to meet the requirements of large skewed workloads common in industry applications. \sys adopts a two-tier record-oriented design to handle larger-than-memory skewed workloads, along with new concurrent latch-free mechanisms and components to maximize performance on modern hardware.
To realize this design, \sys tackles key challenges and introduces several innovations, including new latch-free algorithms for multi-threaded log compaction, a two-level hash index to reduce indexing overhead for cold records, and a read-cache for serving read-hot records.
Our evaluation shows that \sys achieves $2$-$11.9\times$ better throughput compared to existing KV stores, effectively serving the target workload. \sys is open-source and available as part of the \faster project.

\end{abstract}

\maketitle

\pagestyle{\vldbpagestyle}
\begingroup\small\noindent\raggedright\textbf{PVLDB Reference Format:}\\
\vldbauthors. \vldbtitle. PVLDB, \vldbvolume(\vldbissue): \vldbpages, \vldbyear.\\
\href{https://doi.org/\vldbdoi}{doi:\vldbdoi}
\endgroup
\begingroup
\renewcommand\thefootnote{}\footnote{\noindent
This work is licensed under the Creative Commons BY-NC-ND 4.0 International License. Visit \url{https://creativecommons.org/licenses/by-nc-nd/4.0/} to view a copy of this license. For any use beyond those covered by this license, obtain permission by emailing \href{mailto:info@vldb.org}{info@vldb.org}. Copyright is held by the owner/author(s). Publication rights licensed to the VLDB Endowment. \\
\raggedright Proceedings of the VLDB Endowment, Vol. \vldbvolume, No. \vldbissue\ %
ISSN 2150-8097. \\
\href{https://doi.org/\vldbdoi}{doi:\vldbdoi} \\
}\addtocounter{footnote}{-1}\endgroup

\ifdefempty{\vldbavailabilityurl}{}{
\vspace{.3cm}
\begingroup\small\noindent\raggedright\textbf{PVLDB Artifact Availability:}\\
The source code, data, and/or other artifacts have been made available at \url{\vldbavailabilityurl}.
\endgroup
}

\section{Introduction}

Modern large-scale services (search~\cite{chang2008bigtable}, messaging~\cite{rocksdb}, serverless functions~\cite{azure-stream-analytics}) are heavy users of memory and storage. Real-world applications require caches and key-value (KV) stores that offer extremely high throughput at low latencies. Moreover, there is a strong need to deploy such systems in constrained memory environments to reduce the costs of large-scale online services~\cite{zhang2021memory, splinterdb-mapplets}. Many of these services focus on point reads, point updates, and atomic read-modify-writes as their target storage operations~\cite{FASTER}.

Around 2017, the \faster project~\cite{FASTER, faster-demo} was started with the goal of addressing such use cases.
\faster is a key-value storage library that focuses on the problem of achieving bare-metal thread-scalable performance. It was built in C\# along with a port to C++.
Briefly, \faster employs a thread-scalable hash index on top of a hybrid log: a record-oriented storage tier that spans main memory and secondary storage.
\faster was shown to saturate memory bandwidth for in-memory workloads, achieving up to 160M random read operations per second on a single machine. Further, it was shown to saturate disk IOPS for disk-oriented workloads, achieving up to 1M random read operations per second~\cite{FASTER}.
The project was open-sourced in 2018 and has, over the years, seen incredible traction in the research and open-source community, as well as usage in real-world industry scenarios. \faster has around $1$ million downloads on NuGet~\cite{nuget}, $6.5$k stars on GitHub, and over $570$ forks.

We highlight two representative scenarios where the \faster library was integrated into real applications. First, we built a new platform for serverless functions called Netherite~\cite{burckhardt2022netherite, burckhardt2025netherite}, which is a runtime for Microsoft's Azure Durable Functions: a service that allows for the deployment of large-scale stateful serverless applications~\cite{adf}. In Netherite, \faster is used to efficiently store, retrieve, and update the state of individual function invocations. This state is stored across main memory and Azure storage, and the active state is brought back to memory on demand. Second, we integrated \faster into a streaming service for the purpose of saving and retrieving state related to large event records in long-running streaming computations such as temporal joins. Motivated by these use cases, we sought new application scenarios to both reduce existing costs and optimize the \faster design further.

Based on our survey of a variety of services that depend on point-based storage access, we make several key workload observations:

\begin{itemize}[leftmargin=*]
    \item Indexing extremely large state with (1) limited memory and (2) the availability of multiple storage tiers (such as SSD, hard disks, and replicated cloud storage) is a common scenario~\cite{rocksdb-evolution, splinterdb-mapplets, zhang2021memory}.
    \item A natural skew in key access patterns exists for both reads and writes operations (e.g., Zipfian), and the working sets do not entirely fit in-memory (i.e., larger-than-memory)~\cite{rocksdb, rocksdb-evolution, larger-than-memory}.
    \item The \textit{read}-hot and \textit{write}-hot working sets may not fully overlap, necessitating separate treatment for each working set~\cite{rocksdb, twitter-analysis}.
    \item Disk wearing due to excessive writing is a practical concern, due to the large amount of data (i.e., TBs) that is being processed~\cite{disk-wear}.
\end{itemize}

These characteristics apply to use cases such as behavior targeted advertising~\cite{10.14778/1687553.1687590, 10.1109/ICDE.2012.55} in search engines (covered in more detail in Section~\ref{sec:background}), where most tracked users of a search engine---such as those who performed searches in the last 7 days---are inactive at any given moment. They also hold
for use cases such as serverless functions, where the state of most serverless functions
are cold and unused, but we want high performance for the active functions. In streaming systems, we may be tracking billions of records in a temporal join synopsis, but only a small fraction of records may be active (i.e., being joined to new streaming events) at a given time.

The industry’s go-to storage solution for applications that access large skewed workloads has traditionally been Log Structured Merge (LSM) tree-based systems, such as RocksDB~\cite{rocksdb}, which emphasize the judicious use of memory.
This is achieved through a tiered architecture, in which small in-memory components absorb user updates (i.e., memtable) and maintain index metadata (e.g., filters), while large disk components store the actual data (i.e., LSM levels)~\cite{lsm-survey}.
Although this design enables LSM-based systems to store TBs of data, it comes at a high performance cost.
In particular, LSM-based systems deliver main-memory performance far below what modern hardware can achieve~\cite{FASTER, leanstore_evolution}, while their disk-oriented performance is poor, due to their inability to fully utilize available bandwidth of NVMe disk devices (i.e., just $35\%$ NVMe SSD utilization,  Section~\ref{sec:limitations-existing-kvs}).
Other storage solutions such as B-tree based systems~\cite{leanstore, kvell, bw-tree, silo} also do not adequately meet the requirements of these applications, mostly due to their page-oriented design and large write-amplification (i.e., $25$-$90\times$, Section~\ref{sec:limitations-existing-kvs}).

Given the above, a natural question arose: \textit{could we use \faster to improve the throughput of workloads that exhibit the above characteristics?} When we tried to apply the original, unmodified \faster design~\cite{FASTER}, we encountered several practical challenges:

\begin{itemize}[leftmargin=*]
    \item When memory resources are limited, background compaction (or garbage collection) in \faster's single-log design causes increased disk writes (as live records are migrated to the tail), resulting in high disk wear or disruptions in user request processing. It also incurs transient memory spikes as candidate live records need to be tracked in memory during the process.
    
    \item \faster's hash index tracks all live keys in the store and imposes a fixed 8 bytes per-key memory indexing overhead.
    For billions of keys, this leads to a prohibitively large memory footprint.

    \item \faster effectively keeps write-hot records in memory.
    However, when read-hot and write-hot working sets are non-overlapping, read-hot records are either served from disk or brought into the log to be later flushed to disk, incurring additional I/O operations.
\end{itemize}

In this paper, we describe how we have evolved the original \faster C++ design to a new compartmentalized architecture that aims to address all these challenges. The resulting system, \sys (for \faster v2), adopts a two-tier log architecture that inherently handles skewed workloads with greater memory-efficiency, and couples it with high-performance latch-free mechanisms and components that are necessary to saturate the disk bandwidth of modern NVMe storage devices.

However, realizing this design in practice required overcoming key technical challenges.
For instance, ensuring that compacting records across tiers in a CPU- and memory-efficient fashion is critical, while performing compaction concurrently to other user operations (like atomic RMWs) in a safe manner is non-trivial.
To this end, we introduce a \textit{lookup}-based record compaction method that achieves minimal memory and disk I/O overhead, enabling \sys to handle billion-key scale workloads (Section~\ref{sec:tiered-logs}).
This compaction method is based on our new \textit{Conditional-Insert} primitive, which prevents (older) compacted records from overwriting newer versions of the same record (i.e., lost-updates), ensuring overall system correctness, and is multi-threaded, achieving much faster compaction times.
To minimize indexing overhead for cold records, we introduce a concurrent \textit{two-level index} design that spans both memory and disk (Section~\ref{sec:cold-index}).
Finally, we augment a dedicated \textit{read-cache} that provides immediate access to disk-resident read-hot records without any additional I/O overhead (Section~\ref{sec:read-cache}). 

This results in a system that \circled{1} maintains low memory overhead and \circled{2} achieves high performance by exploiting both multi-core CPUs and NVMe storage devices, and \circled{3} remains disk-friendly with minimal write amplification.
We believe \sys is the first comprehensive KV store design to address the practical challenges that limit hash-based storage systems from handling real-world skewed larger-than-memory workloads.

We experimentally evaluate \sys on YCSB and real-world MixGraph~\cite{rocksdb} workloads against several modern key-value stores.
We show that \sys achieves $2$-$11.9\times$ better throughput compared to existing state-of-the-art systems (e.g., original \faster, RocksDB, SplinterDB, KVell, LeanStore), when memory resources are limited.
We also show that \sys matches or outperforms existing solutions even when handling less-skewed workloads, e.g., when $90\%$ of operations access $33\%$ of keys.

\sys is written in C++ as an evolution of the \faster C++ codebase. It is now available in open-source as part of the FASTER project.\footnote{\href{https://github.com/microsoft/FASTER/tree/main/cc}{https://github.com/microsoft/FASTER/tree/main/cc}}.

\section{{\fontsize{10.85pt}{10.85pt}\selectfont Issues with Large Skewed Workloads}}

\label{sec:background}

We start by describing a representative scenario that we identified through discussions with real-world platform builders who use key-value stores, and were considering solutions such as \faster.

\bsubsection{(Targeted Advertising)}
\emph{Search engines perform \emph{behavior targeted advertising}~\cite{10.14778/1687553.1687590, 10.1109/ICDE.2012.55}, for which they store and track per-ad clicks and impressions as well as per-user sketch of ad activity, in a KV store. Queries, to retrieve the sketch for a given user or the number of clicks on a given ad, are point based. Updates are either blind inserts (e.g., insert a new ad into the system) or read-modify-writes (e.g., update the counter or sketch for a given ad or user). The number of users and ads being actively served may be large, with the aggregate data greater than the amount of memory available. For example, users who have interacted with the search engine over a 7-day period may constitute the entire set of tracked users. Further, a long tail of users and ads that are not actively being read or updated, \textit{still} need to be available for immediate queries and updates.
Both the read and write sets exhibit skewness. Finally, the set of users actively browsing and needing a lookup of their sketches (i.e., the read-hot keys) may be different from the ads being frequently shown and having their sketches updated (i.e., the write-hot keys).
}

This application scenario---and the ones described earlier such as serverless functions and streaming---exhibit several interesting workload characteristics.
First, point operations and high throughput are still of paramount importance, and the working sets are \textit{much larger} than main memory~\cite{larger-than-memory, anti-caching}.
Second, the total indexed data is often an order of magnitude larger than memory, with a large fraction of data being \textit{rarely} updated or accessed~\cite{rocksdb, twitter-analysis}.
Third, memory is a \textit{scarce} resource~\cite{zhang2021memory, silt, skimpystash, hotcold-data, splinterdb-mapplets}, periodic memory spikes are not acceptable (as services would have to provision for peak memory), and disk wearing due to excessive writes over the long term is a practical concern~\cite{disk-wear}.
Finally, there is a natural skew in key access patterns for both reads and writes~\cite{skew, twitter-analysis}, but the read and write working sets do not necessarily overlap.

\subsection{Limitations of Existing KV Store Designs}
\label{sec:limitations-existing-kvs}

We observe that existing KV stores, including LSM-based designs and B-tree based ones, do not fully address the requirements of the aforementioned workloads.
We discuss their limitations below.

\begin{figure}[t]
    \centering
    \setlength{\abovecaptionskip}{0mm}
    \includegraphics[width=0.95\linewidth,trim={2cm 5.45cm 0 0.1cm},clip]{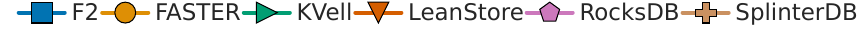}
    \begin{subfigure}[b]{0.44\linewidth}
        \centering
        \includegraphics[width=\textwidth]{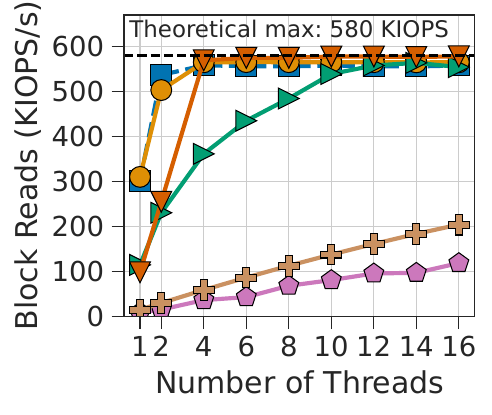}
    \end{subfigure}
    \hspace{0.02\linewidth}
    \begin{subfigure}[b]{0.44\linewidth}
        \centering
        \includegraphics[width=\textwidth]{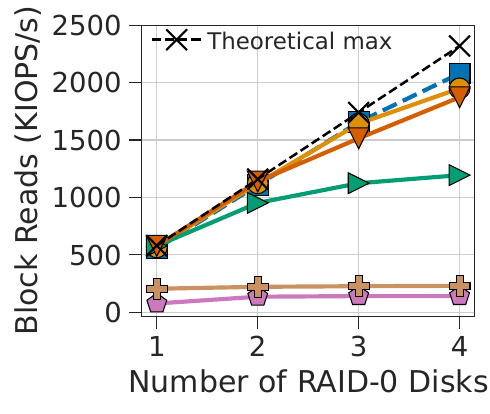}
    \end{subfigure}
    \caption{Out-of-memory performance for uniform read-only workload (30GiB database, 1GiB in-memory buffer), when (a) increasing the number of threads (left), and (b) increase the number of NVMe SSD disks in RAID-0 formation (right).}
    \label{fig:iops-saturation}
    \vspace{-1ex}
\end{figure}

\bsubsection{LSM-Tree-based Designs.}
Log-structured Merge (LSM) Trees~\cite{lsm} designs prioritize memory-efficiency, and they can store TBs of data through their tiered design.
However, they fail to fully utilize the available NVMe SSD bandwidth~\cite{leanstore-nvme}.
To empirically show this, we perform a case study on a system equipped with a 16-core \mbox{Intel Xeon} CPU and four Samsung NVMe PM9A3 SSDs (detailed setup in Section~\ref{sec:evaluation-setup}), with a larger-than-memory (i.e., $30$GiB with $1$GiB in-memory buffer) uniform random read-only workload.
As shown in Figure~\ref{fig:iops-saturation}, we measure random read IOPS ($4$KiB blocks) for five popular key-values stores, when (a) increasing the number of threads used (left plot), and (b) when increasing the number of NVMe SSDs in RAID-0 configuration (right plot).
We observe that both LSM-tree stores, RocksDB and SplinterDB, even when optimized and tuned properly, are not able to saturate I/O bandwidth.

One major factor is the widespread use of filters: given that each point lookup typically accesses numerous such filters, LSMs waste precious CPU cycles that could be used to issue more I/O~\cite{splinterdb-mapplets}.
Further, the benefits of filters can be diminished if the filters no longer fit in memory~\cite{splinterdb-mapplets}.
Prior work has found similar results: they~\cite{leanstore-nvme} have shown that despite existing efforts, LSM stores fail to saturate SSD bandwidth and scale poorly~\cite{FASTER} with increasing number of threads due to write-stalls caused by inefficient data flow across their components~\cite{adoc}.

\bsubsection{B-Tree-based Designs.}
B-Tree based storage designs~\cite{kvell, leanstore, bw-tree} usually rely on in-memory structures (such as index pages, buffer pools, mapping table) to index the keys and cache the respective working set.
Due to their CPU-optimized designs, they can achieve (i) linear thread scalability and (ii) disk IOPS saturation, even for modern NVMe storage devices.

KVell~\cite{kvell} employs a shared-nothing approach, where each thread uses a B-tree index to map keys to a page offset on disk.
When memory resources are abundant, KVell saturates most of the available I/O bandwidth (Figure~\ref{fig:iops-saturation}).
However, when available memory is limited, parts of KVell's index are continuously paged out to disk~\cite{splinterdb},
leading to high read/write-amplification (i.e., $25$--$90\times$) 
and low throughput (i.e., $3$--$10\times$ drop),
as shown in our evaluation (Table~\ref{tab:e2e-ioamp}).
LeanStore~\cite{leanstore, leanstore_evolution} employs a B-tree indexing structure alongside an in-memory buffer manager to support larger-than-memory workloads.
However, due to its page-oriented design, LeanStore performance degrades for skewed workloads.
Specifically, given that hot records can be scattered around all pages, effectively caching the hot set in-memory is not possible.
Perhaps more importantly, we find that the page-oriented design incurs high write-amplification (i.e., $30$-$65\times$),
as for each user update, a disk block write is necessary (Table~\ref{tab:e2e-ioamp}). Bw-Tree~\cite{bw-tree} uses log-structured writes using delta records for pages, and as a result incurs similar compaction overheads as LSM-Tree-based designs for pages with write-hot records.

\section{Original \faster and Limitations}

We now provide background on the original \faster design, including its components and internal mechanisms, before discussing its limitations in handling the above class of applications, motivating the need for evolving the design.

\vspace{-1ex}
\subsection{Design Overview}
\label{sec:faster-background}

\begin{figure}[t]
    \centering
     \includegraphics[width=\linewidth]{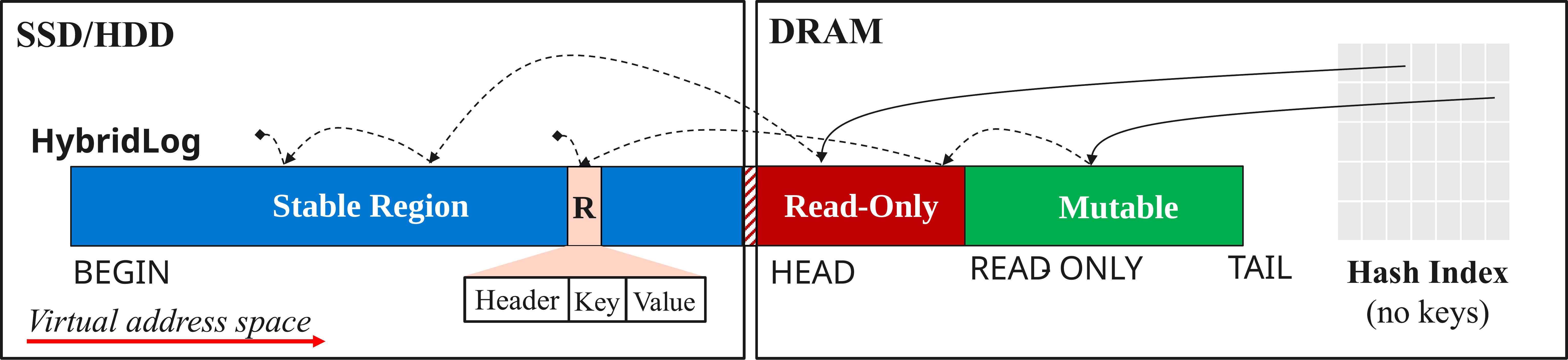}
     \caption{\faster \hlog and index architecture.}
     \label{fig:faster:arch}
     \vspace{-1.5ex}
\end{figure}

\faster~\cite{FASTER} is a log-structured, latch-free key-value store that targets point operations.
As shown in Figure~\ref{fig:faster:arch}, it employs a hash index that chains records stored in a log, which spans both memory and disk (\hlog).
\faster uses a lightweight epoch-based protection framework to facilitate cooperation across threads. 
Due to its log-oriented design, a garbage collection process is invoked periodically to shrink (compact) the log by removing stale tuples.
This is performed using a \textit{scan}-based compaction process that copies \emph{live} records from the beginning of the \hlog to its tail.
As long as both the working record set and the index fit in memory, \faster achieves high performance.
Below, we detail its components.

\bsubsection{Hash Index.}
At its core, \faster consists of a latch-free in-memory hash table, which is divided into cacheline-sized buckets.
Each bucket entry contains a pointer to a record whose key hashes to that bucket.
Each record points to another record, forming a logical linked list of records with common significant key hash bits (i.e., \textit{hash chain}).
Each bucket entry contains additional bits from the associated records' key hash, increasing hashing resolution and further disambiguating what records the bucket entry points without full key comparisons.
A bucket occupies $8$ bytes of in-memory space.
\faster defines four user operations: \texttt{Read}, \texttt{Upsert}, \texttt{RMW}, and \texttt{Delete}.
Latch-free algorithms are used to add/remove entries in the index and to add records at the hash chain tail.

\bsubsection{Hybrid Log.}
Each record pointed to by the hash table is stored in a log that spans disk and main memory, called \hlog. Each record consists of an 8 byte header, a key, and a value. This header, among other information, stores a pointer to the previous address, the log address of the previous record in the hash chain (linked list).
The log itself is divided into three contiguous regions: \circled{1} \textit{mutable}, \circled{2} \textit{read-only}, and \circled{3} \textit{stable} regions.
The mutable and the read-only regions reside in-memory, while the stable one resides on disk.
Records in the mutable region can be atomically updated in-place.
Records in the read-only and stable regions are immutable, and use read-copy-updates (RCU) to the tail, adding to the hash chain tail via a compare-and-swap (CAS) op at the corresponding hash entry, thus providing linearizability guarantees~\cite{linearizability}.

This log design allows write-hot records to be accessed and updated very quickly, while scaling to many threads.
As the tail grows, older log pages need to be flushed to disk and ultimately evicted from main memory.
This is achieved in a latch-free manner by tracking several increasing addresses, i.e., a \texttt{BEGIN} address that tracks the first valid address in \hlog,
a \texttt{TAIL} that tracks the tail of the \hlog address space, etc.

\bsubsection{Epoch Framework.}
\faster uses an epoch-based framework~\cite{FASTER}, which enables synchronization across threads in a lazy fashion, without using fine-grained latches.
Most of the time, threads perform operations independently (e.g., update a record). However, some system-wide events (e.g., flushing log pages to disk) necessitate thread synchronization, i.e., to avoid accessing invalid memory regions or stale data.
This is achieved using a global epoch counter, and thread-local ones, where the latter are periodically synced to the global counter.
This mechanism allows actions to be executed \textit{only after} all threads have agreed to a common view of the world.

\bsubsection{Log Compaction.}
\faster employs a \textit{scan}-based compaction process to perform garbage collection. Here, the oldest disk-resident log records are read from disk and all potentially live records are stored in a temporary in-memory store.
A full scan of the rest of the log confirms which of these records are indeed live. These live records are then inserted into the log tail.
Finally, the \texttt{BEGIN} address is moved forward, effectively truncating the log.

\subsection{Challenges with Large Skewed Workloads}
\label{sec:limitations-faster}

\begin{figure}[t]
    \centering
    \includegraphics[width=0.91\linewidth]{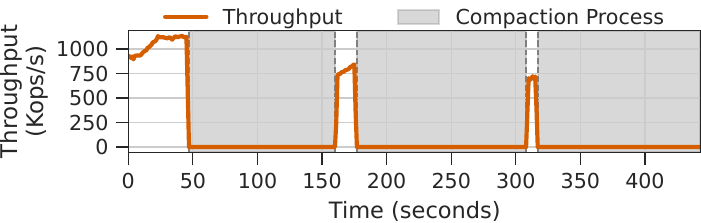}
    \vspace{-1ex}
    \caption{FASTER performance over time for a larger-than-memory RMW YCSB-F workload. Once log size budget is reached, a log compaction process copies cold tuples to the tail of the log. Yet, this results in hot tuples to be evicted to disk (over-and-over), significantly degrading performance.}
    \label{fig:faster-death-spiral}
    \vspace{-2ex}
\end{figure}

\fbsubsection{Single-Log Design Implications.}
The original \faster design uses a single log, which creates new challenges with large skewed workloads.
One such issue is caused by the vast difference in popularity of compacted (cold) vs. in-memory (hot) records.
In particular, during garbage collection (i.e., log compaction), cold records located at the beginning of the log are copied to the log tail.
This not only increases tail contention with incoming user updates (leading to performance degradation), but also causes hot records (stored in-memory) to be flushed to disk, increasing write amplification.
Notice that having cold records occupy the in-memory log region reduces the number of write-hot records that can be in-place updated in the mutable region.
Thus, when log compaction finishes, hot records are appended to the tail of the log, further increasing the overall \hlog size.
This in turn triggers another log compaction process and, as shown in Figure~\ref{fig:faster-death-spiral}, can lead to a "death spiral" behavior, where the system is entirely preoccupied with background compaction operations rather than serving user requests.

Inspired by LSM-tree based designs, \sys addresses these issues by introducing a separate log tier that stores write-cold records, eliminating log tail contention and "death spiral" behavior (Section~\ref{sec:overview}).
\sys also addresses key technical challenges to support efficient, latch-free user operations (e.g., RMWs) in this tiered design (Section~\ref{sec:f2-user-operations}).

\begin{figure*}[t]
    \centering
    \begin{minipage}{0.47\linewidth}
         \centering
         \includegraphics[width=\linewidth]{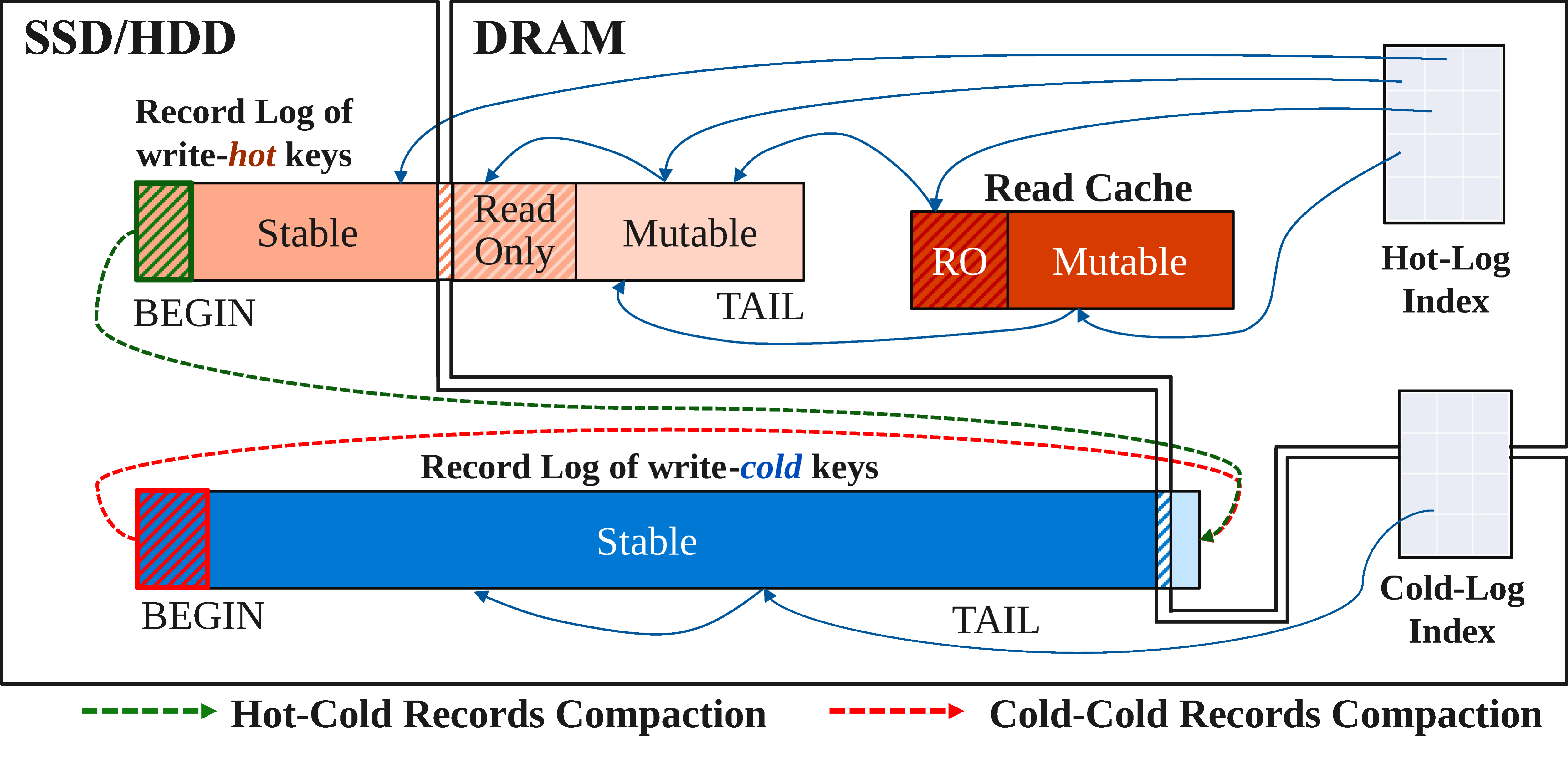}
         \vspace{-3.5ex}
         \caption{\sys Architecture}
         \label{fig:architecture}
    \end{minipage}
    \hspace{0.02\linewidth}
    \begin{minipage}{0.48\linewidth}
         \centering
         \includegraphics[width=\linewidth]{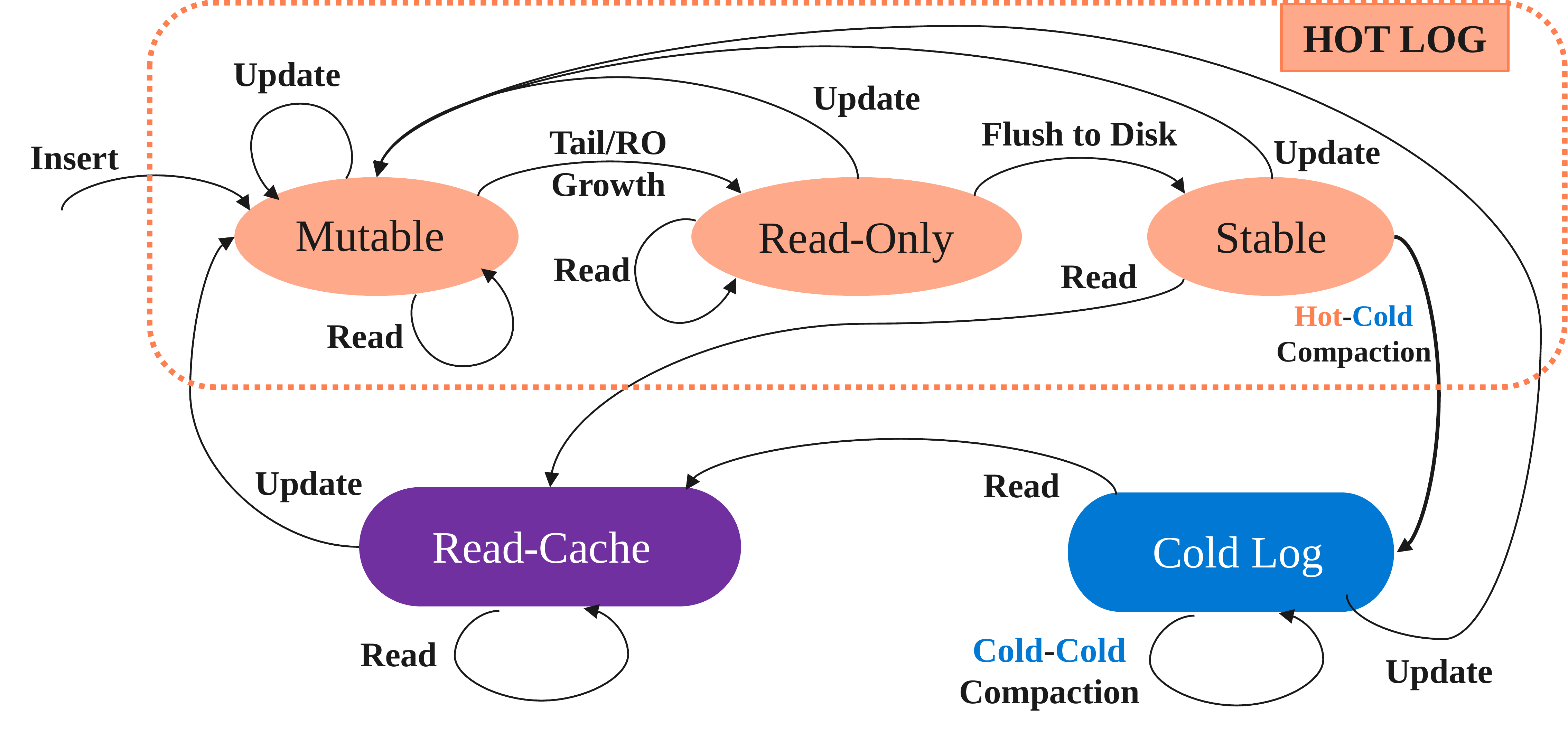}
         \caption{Lifecycle of a Record in \sys}
         \label{fig:lifecycle-record}
    \end{minipage}
    \vspace{-3ex}
\end{figure*}

\bsubsection{Inefficient Record Compaction Mechanism.}
As discussed in Section~\ref{sec:faster-background}, the original \faster design employed a scan-based record compaction process to identify live records and compact them to its log tail.
This scan-based approach has two drawbacks. First, the original implementation was single-threaded, limiting the maximum record compaction rate. Second, it requires (i) additional memory resources (i.e., a temporary memory buffer) to store potentially live records, and (ii) a full scan of the \hlog to reason about record liveness. This approach not only led to transient memory spikes, but also wasted memory and I/O bandwidth resources, which could have been used instead to cache more hot records in-memory and perform more user I/O operations, respectively.

\sys solves both problems by introducing a \textit{lookup}-based compaction mechanism, based on our new \textit{Conditional-Insert} primitive (Section~\ref{sec:conditional-insert}).
This new compaction process is multi-threaded and utilizes only minimal memory and disk bandwidth resources to reason about record liveness (Section~\ref{sec:lookup-compaction}).

\bsubsection{Large Indexing Overhead.}
\faster's hash index tracks all keys in the store and incurs a fixed, per-key memory overhead of 8 bytes.
Although this overhead is manageable for small datasets, when dealing with billions of keys, this leads to a prohibitively large memory footprint (e.g., $64$GiB to index $8$ billion records). 
One might try to constrain the index size, e.g., by restricting the number of buckets, yet, this creates too many hash collisions, making point reads require multiple I/O ops to follow the hash chains on disk.

To address this limitation, \sys introduces a separate two-level hash index that spans both memory and disk, and can index billions of (cold) records with minimal memory footprint (Section~\ref{sec:cold-index}).

\bsubsection{Suboptimal Handling of Read-Hot Records.}
In \faster, write-hot records are effectively in-place-updated in memory through the \hlog design.
However, when read- and write-hot working sets are non-overlapping, read-hot records may get flushed to disk to make space for write-hot records and vice versa. This leads to poor performance: reads incur I/O and writes incur tail growth.

To effectively handle both read- and write-hot working sets, \sys introduces a dedicated in-memory read-cache that provides immediate access to read-hot records (Section~\ref{sec:read-cache}).

\vspace{-1ex}
\section{\sys Overview}
\label{sec:overview}

\sys is a concurrent, lock-free, key-value store that can serve larger-than-memory skewed workloads with \circled{1} low memory footprint, \circled{2} saturation of the available NVMe SSD bandwidth, and \circled{3} minimal disk wear.
\sys supports point \texttt{Reads}, \texttt{Upserts}, \texttt{Deletes} (using tombstone markers), and atomic updates in the form of read-modify-writes (\texttt{RMWs}).
\sys's design also provides linearizable semantics.

Figure~\ref{fig:architecture} depicts \sys's architecture.
First, \sys incorporates a log-structured record store that keeps write-hot keys (i.e., \textit{hot log}), alongside its respective hash index (i.e., \textit{hot-log index}).
Second, it integrates a separate record log for storing write-cold keys (i.e., \textit{cold log}), alongside its respective hash index (i.e., \textit{cold-log index}).
Finally, a \textit{read-cache} lies between the hot-log index and the hot log, which maintains a set of disk-resident read-hot records in a separate in-memory store.
\sys automatically places records on the appropriate component, based on their observed read/write hotness.

\vspace{-1ex}
\subsection{Components Overview}

\fbsubsection{Hot-Log Index.}
The hot-log index employs a lock-free hash table design, similar to the one of the original \faster, that is stored in memory.
Each bucket entry contains a pointer (i.e., address) to a record, whose key hashes to this entry.
However, this record may now reside in either the hot log or read-cache.
Each record points to a (previous) record (if any) in the hot log, forming a hash chain.
Using this hash chain, \sys accesses records with matching key, stored in hot log or in read-cache.

\bsubsection{Hot Log.}
The goal of the hot log is to enable write-hot records to be retrieved and updated promptly, even when many threads are concurrently operating on the log.
Therefore, the hot log is an instance of \faster's \hlog, which is coupled with the  hot-log index.
The hot-log index, indexes only the hot-log (and read-cache) records, requiring fewer memory resources.
Although the organization of the hot-\hlog is largely the same (e.g., insertions at the tail of the log, in-place updates or RCU), we significantly alter its compaction behavior (as we discuss in Section~\ref{sec:tiered-logs}).

\bsubsection{Cold-Log Index.}
The goal of the cold-log index is to reduce the memory resources needed to index cold records.
It is based upon a \textit{two-level} index design, and consists of a (small) in-memory structure and a (large) on-disk one.
The core idea is to group multiple hash index entries together, to create \textit{hash chunks}, and then index these chunks in-memory, while storing the actual chunks on-disk (more details in Section~\ref{sec:cold-index}).

\bsubsection{Cold Log.}
The addition of the cold log enables the physical separation of write-hot and write-cold records.
Cold-log organization is similar to hot-log, with the exception that (almost) all records now reside on disk.
Accessing a cold-log record requires two I/O ops: one for retrieving the hash chain from the cold-log index, and one for reading the actual record from the log.
Yet, the cold-log integration eliminates the contention at the hot-log tail, fixing the "death-spiral" behavior of the original \faster design.

\bsubsection{Read Cache.}
While write-hot records reside in the in-memory part of the hot log, write-cold records do not.
This leads to poor performance for disk-resident \textit{read}-hot records, as \sys has to perform one (or more) I/O operations each time this record is requested.
The read-cache allows immediate access to read-hot records, even if they are stored on disk (more in Section~\ref{sec:read-cache}).

\vspace{-1ex}
\subsection{Lifecycle of a Record}
\label{sec:record-lifecycle}

Figure~\ref{fig:lifecycle-record} depicts the lifecycle of a record in \sys.
The user first inserts a record into the store, by creating a new record in the hot log \textit{tail}.
Initially, the record is created in the mutable region (in-memory), and any subsequent updates are performed in-place. 
As other records are appended to the log, our record eventually moves to the read-only region.
If a user issues an update for this key, we perform an RCU to append the updated record to the hot log tail.
As long as our record is in-memory (read-only or mutable hot-log region), read-cache is not used.

When our record has not been updated for a while, it is eventually flushed to disk (i.e., stable region), as the in-memory regions are populated with newer records.
Here, a user update will result in RCU to the hot-log mutable region, while
a user read will copy the record to the in-memory read-cache (after being fetched from disk).
This allows future user reads to be served directly from read-cache, avoiding any extra I/O operations.
As long as the user issues reads for our record, it remains inside the read cache (i.e., write-cold, read-hot record).
Yet, if no reads occur for some time, our read-cached record is ultimately evicted.

Assuming no further updates, our record ultimately ends up in the back (i.e., \texttt{BEGIN}) of the hot log.
As other records are being appended, the hot log grows larger, which necessitates moving the write-cold records from the hot-cold to the cold-log.
This is achieved through a background \textit{hot-cold compaction} process (green arrow in Figure~\ref{fig:architecture}).
During this process, \textit{live} records from the beginning of the hot log (i.e., compacted region) are copied to the \textit{tail} of the cold log.
Once all live records have been copied, the hot log is truncated, invalidating all records in the compacted region.
Note that during hot-cold compaction, the hot log tail remains intact, and is able to fully accommodate other incoming user requests.

Once the hot-cold compaction finishes, our record now resides in the cold log.
In fact, because the hot set for a skewed workload is relatively small, most records end up in the cold-log.
The cold log resides (almost) entirely on disk, as keeping those (write-cold, read-cold) records in-memory does not bring any benefits.
At this point, a user update request creates a new record to the mutable region of the hot log, while a read request copies our disk-resident record into the read cache (causing it to become read-hot write-cold).
Assuming that no such requests take place, our record remains cold, and eventually arrives at the back (i.e., \texttt{BEGIN}) of the cold-log.

As the cold log is populated with more records, older non-live records need to be garbage-collected.
To do so, we employ another background process, i.e., \textit{cold-cold compaction}.
This process copies \textit{live} cold log-resident records from the back (i.e., \texttt{BEGIN}) of the cold log to \textit{its tail} (red arrow in Figure~\ref{fig:architecture}).
Once all live records have been copied, we truncate the cold log, completely removing non-live records from \sys.
Notice how both hot-cold and cold-cold compactions copy records to the \textit{cold}-log tail, avoiding any tail contention in the hot-log.

\vspace{-1ex}
\subsection{Implementation and Usage}
\label{sec:f2-implementation-usage}

\sys is exposed as an embedded library implemented as part of the \faster C++ code-base~\cite{faster-github}, with around $11$k SLOC. As in the original FASTER implementation, we leverage template meta-programming to avoid runtime overheads, and employ a large set of tests to check for correctness under concurrent execution.

\sys's API for performing user operations is identical to the \faster one, enabling existing users to seamlessly transition to \sys.
Given its more flexible design, \sys provides additional options for users to configuration, to meet the application (and environment) demands.
Listing~\ref{lst:f2-usage} describes how one could configure and initialize \sys.

While optimally tuning the parameters of any KV store is a challenging task~\cite{huynh2021endure}, we provide here some guidelines for configuring \sys.
First, to avoid multiple I/O ops when retrieving disk-resident records, we recommend sizing the hot- (cold-) log indexes based on the expected number of \textit{unique} keys for each log.
For example, indexing $1$B records, with $125$M being hot, requires at least $1$GiB ($16$M buckets $\times$ $64$B each) for the hot-log index, and $256$MiB (i.e., index $28$M hash chunks of $256$B each using $3.5$M buckets) for the cold-log index.
Second, for read-heavy workloads, we recommend trading-off in-memory hot-log space for read-cache one (and vice versa).
For instance, in Section~\ref{sec:eval-f2-impact} we show that properly setting (i.e., increasing) the read-cache size can improve \sys's throughput by $19$-$27\%$.
Finally, the hot/cold  log in-memory organization can be configured based on suggestions from the original \faster paper~\cite{FASTER} (e.g., set log mutable region to $90\%$ of its in-memory size).

The next three sections cover details of each major component in \sys, starting with the tiered record log in Section~\ref{sec:tiered-logs}.

\usemintedstyle{colorful}

\begin{listing}[t]
\centering
\begin{minted}[frame=lines,fontsize=\footnotesize]{cpp}
// Define F2Kv instance
using HotIndex = MemHashIndex<Disk>;
using ColdIndex = ColdIndex<Disk>
using F2KvInst = F2Kv<Key, Value, Disk, HotIndex, ColdIndex>;

// Configuration
ReadCacheConfig rc_config { <mem_hlog_sz> };
F2KvInst::HotIndexConfig hi_config{ <hash_table_sz> };
F2KvInst::ColdIndexConfig ci_config{ <hash_table_sz>, <mem_hlog_sz>};

// Initialize F2Kv
F2KvInst store {
    hi_config, <hot_log_mem_sz>, <hot_log_disk_fp>,
    ci_config, <cold_log_mem_sz>, <cold_log_disk_fp>,
    rc_config };
\end{minted}
\caption{\sys C++ Initialization Code Example}
\label{lst:f2-usage}
\vspace{-1ex}
\end{listing}

\section{Tiered Record Logs}
\label{sec:tiered-logs}

In \sys's tiered design, live records undergo continuous compactions through hot-cold and cold-cold compaction processes.
With hot (and cold) logs potentially managing millions (billion) keys, performing log compaction in a CPU-optimized, memory-efficient manner is of utmost importance.

To this end, we introduce a new primitive, \textit{Conditional-Insert} (\texttt{CI}), that is used as a building block for \sys's lookup-based compaction algorithm and user \texttt{RMW}.
Then, we describe how threads perform record compaction and user operations, emphasizing on correctness issues under concurrent execution.
Finally, we highlight such an issue that arises when \texttt{Reads} are performed concurrently with cold-cold compaction, and explain how it is addressed.

\subsection{Conditional-Insert Primitive}
\label{sec:conditional-insert}

\begin{figure}
    \centering
    \includegraphics[width=.95\linewidth]{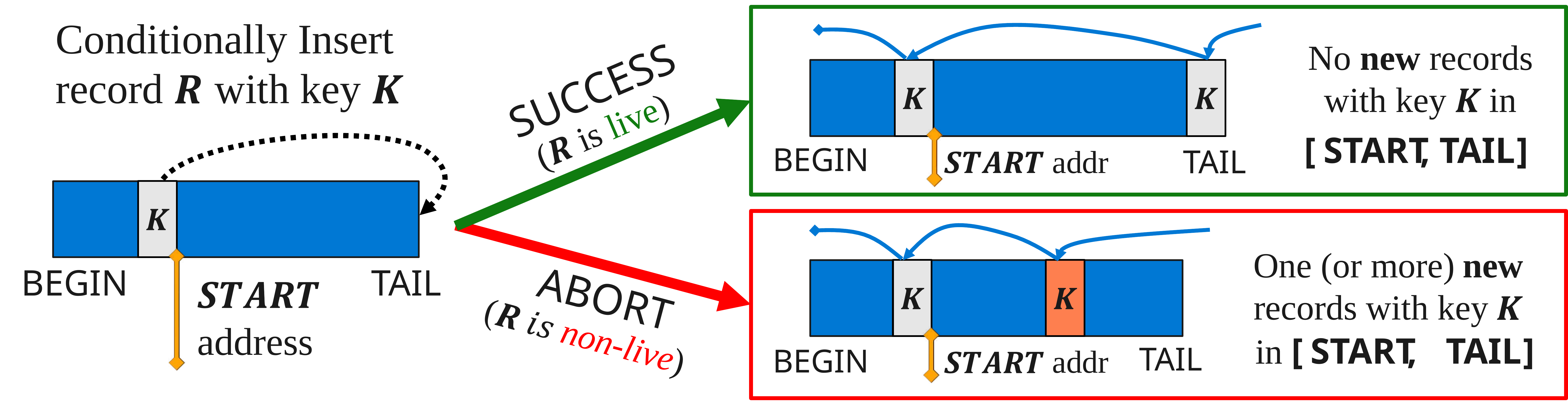}
    \caption{\textit{Conditional-Insert} op and possible outcomes.}
    \label{fig:conditional-insert}
    \vspace{-2.5ex}
\end{figure}

Our goal is to develop a primitive that can append a record to a log, only if no other record with a matching key has been appended in the meantime.
In other words, we want to insert this key \textit{conditional} on no newer insert happening during the process.
More formally, given a record $R$, stored in a record log (i.e., \textit{source} log) and a \texttt{START} address in the log, \texttt{CI} appends the record to the tail of \textit{target} record log (same or different to the source log), only if there exist no record(s) with a matching key in the \texttt{(START}, \texttt{TAIL]} address range of the source log.
If a record exists in this range, the operation aborts (i.e., becomes a no-op).
Figure~\ref{fig:conditional-insert} depicts these two possible outcomes (i.e., success, abort).
For ease of exposition, we initially make two assumptions: (1) the source log is the same as the target log, and (2) the \texttt{START} address matches the log address of $R$ we wish to append to log tail.

Conditional-Insert is implemented as follows.
First, we perform a lookup at the index of the source log, to find the entry corresponding to $R$'s key, and we store a copy of that \texttt{entry} in the operation context.
The index entry contains the log address of the most-recent record (of this hash chain) in the log.
Starting from this address, we follow this hash chain backwards, possibly issuing read I/O request(s).
If at any point during this backwards search, we encounter a record that matches our key, we promptly \textit{abort} (i.e., \textit{non}-live record).
Otherwise, we arrive either at the end of the hash chain or at some address outside the search range (i.e., \texttt{address} $<$ \texttt{START}), meaning we can now copy $R$ to the target log tail.

Appending $R$ to the log tail must now ensure that no other newer records with the same key were inserted in the meantime (e.g., by a concurrent user or compaction operations).
Otherwise, it is possible that we accidentally overwrite a newer version with an older one (i.e., lost update anomaly).
Hence, we leverage the previously saved index \texttt{entry}.
In particular, we first write the record to the log, and then perform an index update (using an atomic CAS operation), expecting that the index \texttt{entry} remained unchanged.
If CAS fails, then newer records were inserted in this hash chain.
In this case, we invalidate our written log record, and restart our search on the hash chain, but \textit{only} check the newly-introduced records on the hash chain. 
As before, if we encounter a (newer) record with a matching key, we abort; otherwise we try to update the index entry again.
We repeat the above process, until either succeeding at appending the record (i.e., record is live), or aborting altogether.
This ensures that in either case, no newer record(s) have been overwritten.

\vspace{-1.5ex}
\subsection{Lookup-based Compaction}
\label{sec:lookup-compaction}

We now describe how we can leverage the \texttt{CI} primitive to make compactions faster and more memory-efficient.
Given a source log and a (potentially different) target log, lookup-based compaction process consists of the following phases:

\noindent
\circled{1} \textbf{Copying phase}: Starting from the beginning of the source log (oldest records), we sequentially scan a fixed \texttt{[BEGIN, UNTIL]} range.
This range represents a specific percentage of the entire log (e.g., $10\%$).
For each record scanned, we issue a Conditional-Insert operation.
If the record is live, then a copy is atomically created at the tail of the target log.
Note that during this phase, copies of the \textit{same} live records may exist in both logs (for hot-cold compaction) or in both ends of the cold log (for cold-cold compaction).

\noindent
\circled{2} \textbf{Truncation phase}: Once we process all keys in this source log region (i.e., all live records have been copied to target log), we \textit{truncate} the source log, by atomically setting the \texttt{BEGIN} address to \texttt{UNTIL}.
Then, all hash index entries that point to invalid addresses (i.e., \texttt{address$~<~$BEGIN}), are invalidated using CAS operations.

\bsubsection{Cold-Cold Compaction.}
Our initial assumptions with \texttt{CI} were: (1) the source log is the same as the target log, and (2) the \texttt{START} address is the log address of the record we wish to append to the log tail.
Notice how with these assumptions, \texttt{CI} can safely compact a \textit{live} cold-log record to the cold-log tail, even when newer records are being appended to the same tail (e.g., hot-cold compaction).

\bsubsection{Hot-Cold Compaction.}
Here we consider the case where the source log (i.e., hot log) differs from the target log (i.e., cold log), relaxing our first assumption.
As before, we follow the hash chain backwards for any key matches, ultimately exploring the entire hot log address range.
We are now ready to copy our record to the cold log tail.
Since the records in the cold log are older by-design, they naturally satisfy our invariant and we can just issue an \texttt{Upsert} to cold log.
The only implication here is that we might \texttt{Upsert} non-live keys (e.g., if newer records entered the hot log in the meantime).
While these superfluous writes might lead to slightly more disk operations, correctness is still ensured.
We late relax our second assumption, when discussing user \texttt{RMW}s (Section~\ref{sec:f2-user-operations}).

\fbsubsection{Concurrent Conditional-Insert.}
Our key invariant is satisfied, even in the presence of concurrent \texttt{CI} ops.
The only concern here is a possible record re-ordering at the target log (as a result of non-sequential record processing), that might lead to overwriting of newer records.
First, we note that when compaction threads operate on records with different keys, records re-ordering poses no correctness issues (i.e., different hash chains).
Now consider a scenario where two threads, $T_1$ and $T_2$, operate on two different records with the same key, $R_1$ and $R_2$.
We know that both records are part of the same hash chain, and thus one record, e.g., $R_1$, is located in front (i.e., higher log address) of the other, e.g., $R_2$, in the hash chain.
This suggests that $R_1$ is live, while $R_2$ is not.
When the two threads call \texttt{CI}, \textit{only} $T_1$'s request succeeds.
This is because by following the hash chain backwards, $T_2$ inevitably encounters $R_1$ (i.e., same key) and thus aborts.
Notice how $T_1$ does not encounter $R_2$ at all, since it is located before $R_1$ in the log.
Generalizing this to many threads operating on multiple records with the same key, it follows that \textit{exactly one} record for each key is compacted.

\bsubsection{Multi-threaded Compaction.}
To achieve shorter compaction times, multiple threads can participate in the compaction process, issuing concurrent \texttt{CI} ops on different records.
Participating threads coordinate using the epoch protection framework (overhead is negligible), and correctness under concurrent \texttt{CI} ops is always ensured (see above discussion).
During copying phase, we employ an in-memory circular buffer that is populated with several ($32$MiB) log pages; initially the first pages in the \texttt{[START}, \texttt{END]} range.
Records residing inside the log pages are distributed to compaction threads using fetch-and-add atomic operations, while the next log page(s) (if any) are prefetched from the disk to avoid any processing stalls.

\bsubsection{Summary.}
Unlike \faster's scan-based compaction, \sys's lookup-based compaction (i) requires minimal memory resources (three log pages, or $96$MiB), (ii) is multi-threaded, enabling much faster compaction times (i.e., $5.2\times$ as shown in Section~\ref{sec:eval-f2-impact}), and (iii) performs only the absolute necessary disk operations to determine record liveness.
Hence, \sys can even compact billion-key logs, which would otherwise be infeasible with the original \faster.

\vspace{-.5ex}
\subsection{User Operations}
\label{sec:f2-user-operations}

\fbsubsection{Upsert and Delete.}
We implement \texttt{Upserts}/\texttt{Deletes} as follows.
First, we perform a lookup in hot-log index, then append the new record to hot log tail, and finally CAS the index entry to point to the our newly-appended record.
If CAS fails, we mark the record as invalid and retry.
In \texttt{Delete}, a tombstone record is \textit{always} inserted, even when the entry for the key does not exist in the hot-log index, as (non-tombstone) valid records may still exist in the cold log.

\bsubsection{Read.}
We first issue a \texttt{Read} op in the hot log (i.e., most recent records).
If a record is found there, we return it to the user.
If a tombstone record is found, we return \texttt{NOT\_FOUND}.
If no record is found in the hot log, we then issue a \texttt{Read} to the cold log.
As before, we either return a valid record, or \texttt{NOT\_FOUND}.
The above algorithm provides correct results in most cases.
Yet, under concurrent cold-cold compaction, a \texttt{Read} might return \texttt{NOT\_FOUND} even if a record exists.
We discuss this anomaly in Section~\ref{sec:false-absense-anomaly}.
We later explain how \texttt{Reads} are modified when using a read-cache (Section~\ref{sec:read-cache}).

\setlength{\textfloatsep}{1ex}

\begin{algorithm}[t]
\footnotesize
\SetKwProg{Fn}{function}{:}{}
\SetKwFunction{FnRmw}{Rmw}


\Fn{Status \FnRmw{key}}{
    \SetKwData{StartAddress}{start\_addr}
    \SetKwData{HotLog}{hot\_log}
    \SetKwData{ColdLog}{cold\_log}
    \SetKwData{Index}{index}
    \SetKw{Goto}{goto}
    
    \SetKwData{ReadStatus}{read\_status}
    \SetKwData{RmwStatus}{rmw\_status}
    \SetKwData{CiStatus}{ci\_status}
    \SetKwData{Status}{\color{BrickRed}Status}
    \SetKwData{StatusNotFound}{\color{BrickRed} NOT\_FOUND}
    \SetKwData{StatusOk}{\color{BrickRed} OK}

    \SetKwData{Retry}{\bf \color{Sepia} RETRY}
    
    \SetKwData{NewValue}{new\_value}
    \SetKwData{Address}{address}
    
    \SetKwFunction{FindEntry}{FindEntry}
    \SetKwFunction{Rmw}{Rmw}
    \SetKwFunction{Read}{Read}
    \SetKwFunction{ConditionalInsert}{ConditionalInsert}
    \SetKwFunction{UpdateValue}{UpdateValue}
    \SetKwFunction{InitialValue}{InitialValue}
    
    \StartAddress = \HotLog.\Index.\FindEntry{key}.\texttt{address}\;

    {\color{blue}\tcp{Try RMW record in hot log}}
    \RmwStatus = \HotLog.\Rmw{key, create\_if\_not\_exists=false}\;
    \If{\RmwStatus $\neq$ \Status.\StatusNotFound}{
        \KwRet{\RmwStatus}  ~~{\color{blue} \tcp{Record updated!}}
    }
    
    {\color{blue} \tcp{No record in hot log -- try Read from cold log}}
    \ReadStatus = \ColdLog.\Read{key, record}\;

    \If{\ReadStatus == \Status.\StatusOk} {
        \NewValue = \UpdateValue{key, input, record.value}\;
    } \lElse{
        \NewValue = \InitialValue{key, input}
    }

    {\color{blue}\tcp{Check if stored start log address is valid}}
    \If{\StartAddress $<$ \HotLog.\texttt{begin\_address}}{
        \Goto \Retry\;
    }
    
    {\color{blue} \tcp{Append updated value; abort if new records}}
    \CiStatus = \HotLog.\ConditionalInsert{key, new\_value, \StartAddress}\;
    \If{\CiStatus == \Status.\StatusOk}{
        \KwRet{\Status.\StatusOk}\;
    } 
    \BlankLine
    \Retry:\\
    \KwRet{\Rmw{key}}\;
 }
\caption{User Read-Modify-Write (\texttt{RMW}) in \sys}
\label{lst:rmw-algo}
\end{algorithm}

\bsubsection{Read-Modify-Write.}
A user RMW operation \textit{atomically} updates the value of a key based on user-provided logic, or inserts a record with an initial value if the key does not exist.
The first step in \texttt{RMW} is to locate the most-recent record with a matching key. In \sys, this record may reside in either hot or cold log.

Algorithm~\ref{lst:rmw-algo} details how user \texttt{RMWs} are performed.
First, we issue a \texttt{RMW} request to the hot log (L3).
Since write-hot records are usually stored in the hot log, this makes the common case fast, as we can quickly return upon updating (possibly in-place) the record (L5).
If no record matching our key exists in hot log, we refrain from creating a new record.
Instead, we issue a \texttt{Read} request to the cold log (L6), as a record may exist there.
In this case, we update its value (L8) using the user-provided logic (i.e., \texttt{UpdateValue}); otherwise (L9) we use the initial value (i.e., \texttt{InitialValue}).
Finally, we try to append the updated record to the hot log tail (L12).

Concurrent to our RMW operation, newer records with the same key might have been appended by the user, causing a change in the hash chain.
To ensure that we always use the most-recent record for our key, we leverage the \texttt{CI} primitive we presented earlier.
More specifically, at the very start of the user \texttt{RMW} (L2) we fetch and store the address where the hash chain begins in the log (i.e., \texttt{start\_addr}).
We later use that address to determine whether any new records has been inserted in the \texttt{(start\_addr,}\texttt{TAIL]} range (L12).
If this is the case, we abort the user RMW operation and retry again.
Note that the hot log RMW request (L3) will now most likely succeed, since a record now exists in the hot log (assuming small chance of hash collisions).
Otherwise, \texttt{CI} successfully inserts the updated record to the tail of hot log.
In the rare case where the range \texttt{(start\_addr,TAIL]} is invalid (L10), caused by log truncation (e.g., due to concurrent hot-cold compaction), we retry from the start.

\subsection{False-Absence Anomaly}
\label{sec:false-absense-anomaly}

A cold log \texttt{Read} traverses the entire log, by following the hash chain.
However, it is possible to fail locating a record that is indeed present in the log, incorrectly returning \texttt{NOT\_FOUND}.

\begin{figure}[t]
    \centering
    \includegraphics[width=\linewidth]{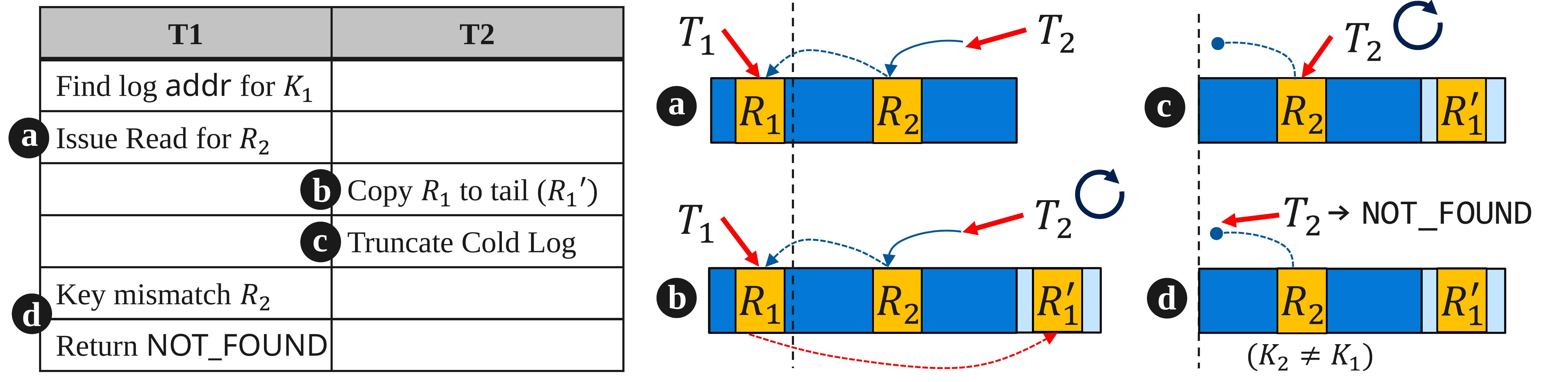}
    \caption{False-Absence Anomaly Scenario}
    \label{fig:false-absense}
\end{figure}

Consider the scenario depicted in Figure~\ref{fig:false-absense}, where a \texttt{Read} operation is issued in the cold log (after a failed search in the hot log) for a given key, $K_1$.
At the same time a concurrent cold-cold compaction is being executed.
Assume that only a single record $R_1$ for key $K_1$ exists in the cold log, and it is located in at the very beginning of the cold log.
The following events transpire in-order.
First, thread $T_1$ issues a \texttt{Read}, and thus performs a lookup in the cold-log index to find the log address of the first record in the hash chain (i.e., $R_2$).
$T_1$ then issues a read I/O request to fetch $R_2$ from disk.
Unbeknownst to $T_1$ at the time, $R_2$ has a different key $K_2 \neq K_1$, i.e., due to a hash collision.
While $R_2$ is being read from disk, thread $T_2$ performing the compaction, manages to copy all live records to the tail of the cold log.
Thus, a copy of $R_1$, $R_1'$ has been written to the tail.
$T_2$ then proceeds and truncates the log, invalidating $R_1$.
Then, $R_2$ is finally fetched from disk.
$T_1$ only now realizes that $R_2$'s key $K_2 \neq K_1$.
Now, it follows the hash chain backwards, only to find that the previous address, originally pointing to $R_1$, is now invalid (due to log truncation).
Thus, $T_1$'s \texttt{Read} op returns \texttt{NOT\_FOUND}, as it deduces that no record with key $K_1$ exists in either hot or cold log.
However, this is clearly incorrect, as $R_1'$ exists in the cold log tail.

This anomaly occurs because $T_1$ is not aware of the concurrent cold-cold compaction.
To address this, one might try to employ some locking scheme (e.g., where cold log truncation is done only when no cold-log \texttt{Reads} are active), or temporarily store every live record in compacted region into a separate in-memory store.
However, the former introduces starvation issues (e.g., for constant stream of cold-log requests), while the latter introduces additional memory overhead.
Instead, we fix this issue by employing a shared atomic counter that tracks the number of completed log truncations.
On every cold-log \texttt{Read} request, we now first fetch and store (in the operation context) the number of log truncations.
We then follow the respective hash chain backwards, as before.
If we ultimately find no record with a matching key, we then check if a log truncation took place, i.e., by comparing the current counter value with the one we stored previously.
If a log truncation indeed occurred, we traverse just the \textit{newly}-introduced part of the hash chain (if any) to check whether a record was indeed compacted.
While this scheme introduces some extra work to few cold-log \texttt{Read} ops to ensure correctness, log truncations are infrequent, thus the common case remains unaffected.

\vspace{-1ex}
\section{Indexing Cold Records}
\label{sec:cold-index}

When \sys handles larger-than-memory skewed workloads, most records end up in the cold log.
Yet, indexing this many records using solely in-memory structures incurs large memory overheads.
For instance, indexing a billion keys using a design similar to the hot-log index requires at least $8$GiB.
Other systems require even more memory, as they store extra metadata (e.g., $19$GiB for KVell).

\begin{figure}[t]
    \centering
    \includegraphics[width=0.92\linewidth]{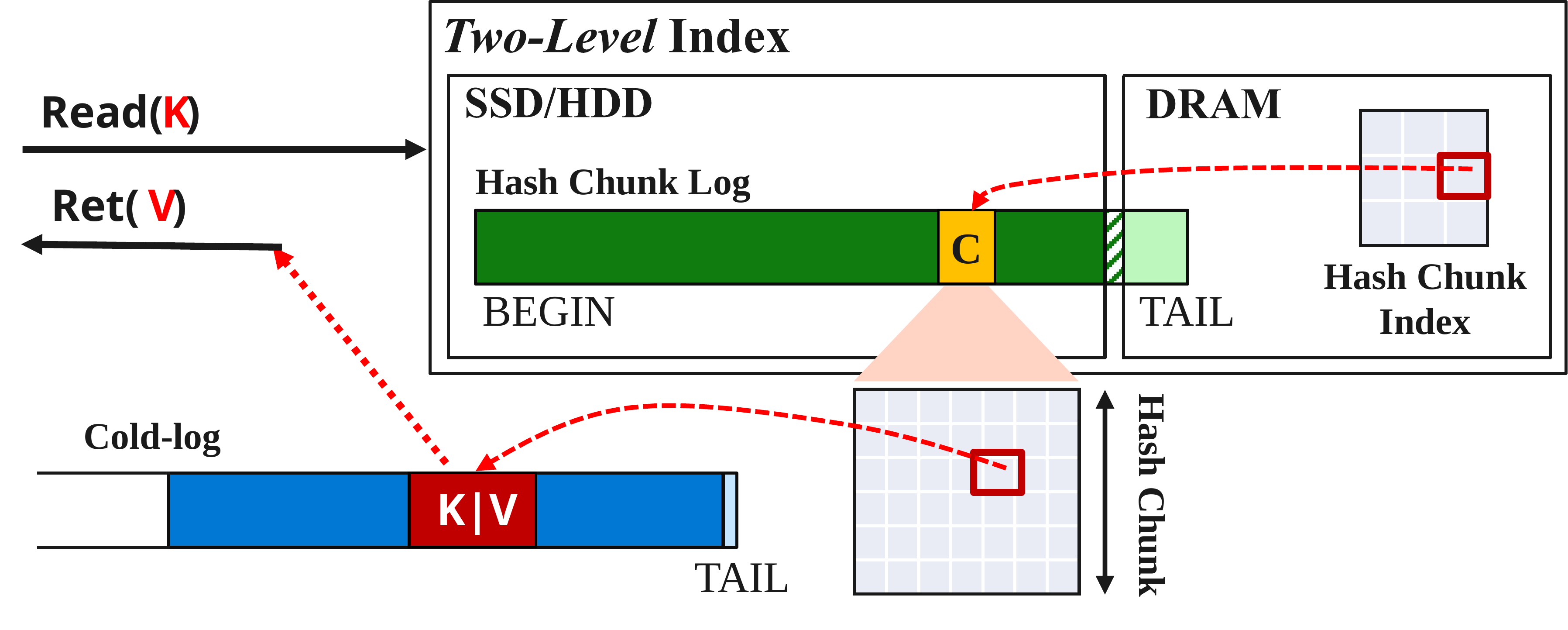}
    \vspace{-1.25ex}
    \caption{Cold-log Index architecture, and \texttt{Read} op lifecycle.}
    \label{fig:two-level-index}
\end{figure}

To this end, we introduce a \textit{two-level} hash index design, as shown in Figure~\ref{fig:two-level-index}.
The core idea is to perform in-memory indexing at a coarse-grained level.
Specifically, we first group multiple hash index entries together, to create \textit{hash chunks}.
Each hash chunk holds a fixed power-of-two number of entries (e.g., $32$).
Then, we use an in-memory hash table to index these chunks ($1^{\text{st}}$ level), while the actual chunks are stored in a log-structure disk store ($2^\text{nd}$ level).
To facilitate concurrent updates on the index chunks, we leverage a \hlog instance, configured with a small in-memory region.

This cold-log index design comes with three main benefits.
\circled{1} By reusing existing lock-free components (i.e., \hlog), we can easily ensure multi-threaded correctness.
\circled{2} The hash chunk size allows us to control the in-memory overhead of the index: increasing (decreasing) hash chunk size results in fewer (more) hash chunks required to index the same number of keys, leading to a smaller (larger) in-memory hash table.
Finally, \circled{3} by adopting a log-structure design to store hash chunks, we can now use small chunks (i.e., less than $4$KiB disk block size), leading to low write amplification, as a single disk block write updates many chunks. We note that our cold-log index design could be replaced by any other concurrent memory-efficient index structure as well.

\bsubsection{Finding an Entry.}
Finding a cold-log hash entry consists of (i) reading the hash chunk from the hash chunk log, and (ii) extracting the respective hash entry.
Given a key, we first compute its hash value, $h$, and then identify the respective hash chunk using a subset of $h$'s bits.
Then, we issue a \texttt{Read} in the hash chunk log.
Once the chunk is read, we identify the hash entry for our key (different subset of $h$'s bits).
Finally, we issue a \texttt{Read} in the cold-log, as this hash entry now points to a cold-log hash chain.

\bsubsection{Modifying an Entry.}
Modifying a cold-log hash entry consists of (i) reading the hash chunk from the hash chunk log, (ii) applying our update to the specific hash entry, and (iii) writing the entire hash chunk back to the hash chunk log.
Notice how this process naturally fits the functionality of a RMW.
Thus, given a record key, we first identify the respective hash chunk (from the key hash), and then issue an \textit{atomic} RMW to the hash chunk log; if no hash chunk exists, we create an empty one (i.e., all entries are invalid).

\bsubsection{Configuration.}
We configure the index so that we use less than $1$B of memory per cold key.
We set the size of each hash chunk to $256$B, resulting in $32$ hash index entries being stored in a single chunk.
Smaller chunk sizes (e.g. $64$B) trade off less write amplification for a larger index memory footprint, and vice-versa.
We set the number of chunks based on the expected number of unique cold keys, e.g., for $250$M keys, we can use $250\text{M}/32 \approx 8\text{M}$ chunks (i.e., one hash entry per key).
Indexing $8$M chunks in-memory requires just $64$MiB.

\vspace{-1ex}
\section{Serving On-Disk Read-Hot Records}
\label{sec:read-cache}

\sys's two-tier design physically separates write-hot records from write-cold ones.
However, this leads to poor performance when serving disk-resident read-hot records, as \sys would have to perform I/O each time such records are requested.
To address this, we introduce a dedicated, in-memory read cache.

We tightly integrate the read-cache with the hot-log index, by extending the hot-log index chains to span both the read-cache and the hot-log (Figure~\ref{fig:architecture}).
Specifically, a hot-log index entry may now (optionally) point directly to a record in the read-cache, before continuing to point one (or more) hot-log records.
The entry may also point to the hot-log directly, bypassing the read-cache altogether.
Read-cache keeps \textit{replicas} of records originally located in the hot or cold log;
the original records are never removed from their position in the log.
Records already residing in-memory (e.g., hot-log in-memory region) are never inserted in the read-cache.

Read-cache is implemented as a separate in-memory \hlog, and thus contains only a mutable and a read-only (RO) region.
Records are inserted at the log tail (i.e., mutable region), and eventually evicted at the head (i.e., RO region).
If a record residing in the RO region is requested again by the user, we update its presence in the read-cache, by copying it to the log tail.
This gives our record a second-chance (similar to Second-Change FIFO cache~\cite{tanenbaum-os}), and further ensures that the most read-hot records are never evicted.

The key invariant that the read-cache should satisfy is that for a given key, it should keep its \textit{most-recent} record.
We next describe how user operations are modified to ensure this.

\bsubsection{Upsert, RMW, Delete.}
When inserting/updating a record for a given key to \sys's hot log, we need to invalidate the (older) read cache-resident entry (if any).
Since the hot-log and read-cache share the same hash chains, we can do so by traversing the record hash chain, invalidating any potential records located in the read cache (by setting a bit in the record header), and then proceed with the request as before (Section~\ref{sec:f2-user-operations}).

\bsubsection{Reads}
Given a key, we first identify its respective hash chain (by consulting the hot-log index), and then follow the chain backwards as before.
If we encounter a valid record with a matching key in read-cache (i.e., most-recent), we promptly return it to the user.
Otherwise, we continue the search in the hot (and possibly in the cold log).
If we find a disk-resident record in either log, we first try to insert it to the read-cache; this might fail if another thread concurrently updates the same key.
Ultimately, we either return a valid record to the user or \texttt{NOT\_FOUND}.

\bsubsection{Records Eviction}
When the read-cache becomes full, records need to be evicted to make space for other (read-hotter) records.
The only challenge is to ensure that both during, and after the eviction process, the system always remains in a consistent state. 
Thus, the records eviction process involves re-arranging hash chains that point to soon-to-be-evicted read-cache records, in a lock-free manner, without impeding incoming user ops.

Eviction is performed on a per-page granularity.
For each record in the page, we first determine if it is valid; if it is, we skip the record.
Otherwise, it means that the respective hash index entry points to this soon-to-be-evicted record, and thus needs to be modified to point to the hot log instead.
To do so, we retrieve the next address in the hash chain, which always points to the hot log (i.e., at most one record per hash chain resides in read cache).
Then, we CAS this address to the hash bucket entry; 
if we succeed, we move to the next record.
Otherwise, some other thread altered the hash chain in the meantime, e.g., by inserting a new record in the chain.
Therefore, our record has been invalidated, and we again move to the next record.
Note that multiple threads can participate in this process safely, as no two evicted records share the same hash chain.

\section{Evaluation}
\label{sec:evaluation}

\sys was recently merged into the \faster code-base, and we do not yet have production workloads to test it on. However, we find that we can emulate the characteristics of larger-than-memory skewed workloads with available YCSB and MixGraph benchmarks.
In this section, we evaluate \sys along multiple dimensions and and compare against state-of-the-art KV stores. 
In particular, we show that:

\begin{vinlist}
    \item \sys's lookup-based compaction is $5.2\times$ faster than \faster's scan-based one and uses $25\times$ less memory. \sys's read-cache improves throughput by up to $1.27\times$ for read-heavy workloads (Section~\ref{sec:eval-f2-impact}).
    \item \sys provides meaningful speedups ($2$--$11.9\times$) across YCSB and real-world MixGraph workloads, with on average $1.4$-$2.9\times$ lower write amplification than LSM stores (Section~\ref{sec:eval-baselines}).
    \item \sys remains robust to varying degrees of key access skewness, outperforming the best SoTA stores by on average $2.6\times$ ($1.8\times$) for high (low) skewness levels (Section~\ref{sec:eval-skewness}).
    \item \sys outperforms the best systems on small memory budgets ($2.5$-$10\%$ of DB size) by $2.7\times$ on average, while matching the best system for larger ($\ge20\%$ of DB size) budgets (Section~\ref{sec:eval-mem-budget}).
\end{vinlist}

\vspace{-1ex}
\subsection{Experimental Setup}
\label{sec:evaluation-setup}

\begin{figure}[t]
    \centering
    \includegraphics[width=0.93\linewidth]{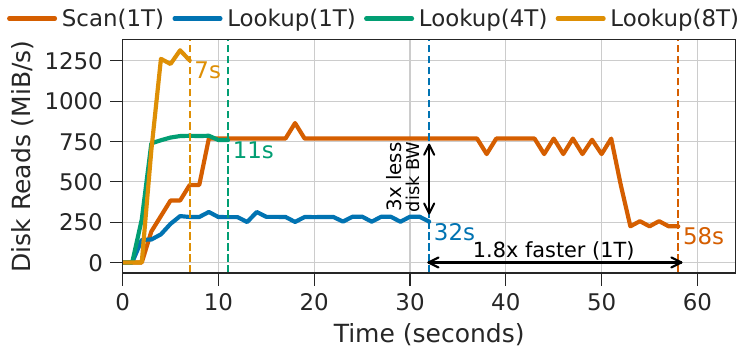}
    \caption{\faster's scan-based vs \sys's lookup-based single-log compaction disk read throughput, when compacting 2GiB (out of 30GiB). For same target disk BW, lookup-based finishes 5.2$\times$ faster and uses $25\times$ less memory ($120$MiB vs $3$GiB).}
    \label{fig:lookup-vs-scan-compaction}
\end{figure}

\fbsubsection{System.}
We conduct all experiments on CloudLab~\cite{Cloudlab}, using a \texttt{sm110p} node that is equipped with a $16$-core Intel Xeon Silver 4314 CPU, $128$GiB of RAM, and four Samsung PM9A3 NVMe SSD (PCIe v4.0) devices, running Linux kernel v$5.4$.
For our experiments, we use all four NVMe disks in RAID-0 formation (using \texttt{mdadm}) with \texttt{ext4} filesystem and disk block size of $4$KiB.

\bsubsection{YCSB.}
We use YCSB~\cite{ycsb} workloads with $250$ million keys ($8$B keys, $100$B values) and run YCSB-A ($50\%$ reads, $50\%$ updates), YCSB-B ($95\%$ reads, $5\%$ updates), YCSB-C, ($100\%$ reads), YCSB-F, ($50\%$ reads, $50\%$ RMWs), YCSB-D read-latest workload ($95\%$ reads, $5\%$ inserts), and a custom YCSB-W ($5\%$ reads, $95\%$ updates).
We model the skewness of real-world workloads with a \textit{Zipf} distribution, using the \textit{YCSB default} $\theta=0.99$ or equivalently $\alpha=1/(1-\theta)=100$, and other skewness factors ($\alpha \in [3, 1000]$).
With the default YCSB skewness factor (\mbox{$\theta=0.99$ / $\alpha=100$}), $90\%$ of accesses go to $18\%$ of records.

\bsubsection{MixGraph.}
We employ two MixGraph (MG) benchmarks with $250$M keys, i.e., All-Dist (AD) and Prefix-Dist (PD), which were developed to emulate the real-world workloads observed in Meta's production services~\cite{rocksdb}.
They use $48$B keys with variable-sized values, and consist of $83\%$ reads, $14\%$ writes (and $3\%$ seeks, which we skip).
All-Dist clusters hot keys close together in the key-space, while Prefix-Dist further partitions the key-space into smaller hot/cold key-ranges and issues more requests to hot ranges.

\bsubsection{Measurements \& Resources.}
For each experiment, we load the dataset into the system (e.g., $30$GiB), warm it up with $25$M ops, and then run each workload for $300$M ops measuring system throughput.
We report average throughput in thousands of requests per second.
Unless otherwise noted, we set the available memory to $10\%$ of the dataset set, i.e., $3$GiB for YCSB $250$M key-value dataset, and $4$GiB for MixGraph one.
We also manually pin user threads to hardware cores.
The above is done via the exposed user configuration parameters of each system and further enforced via Linux \texttt{cgroup}~\cite{cgroup}.
This setup is similar to prior works~\cite{splinterdb, treeline, rocksdb}.

\bsubsection{Baseline Systems}
We compare \sys against several state-of-the-art systems including, SplinterDB~\cite{splinterdb, splinterdb-mapplets} (commit \texttt{1939a12}), RocksDB v8.11.4~\cite{rocksdb}, FASTER~\cite{FASTER}, Kvell~\cite{kvell} (commit \texttt{af10b7a}), and LeanStore \cite{leanstore} (commit \texttt{26d4a46}, \texttt{io} branch).
We configure all systems to use Direct I/O disk ops, and disable any persistent layer (e.g., write-ahead logging), compression, and checksums (if supported).

For all baselines, we set their memory-related parameters as recommended by their documentations (for the given memory budget), and apply all available point operation optimizations.
For RocksDB, we enable Bloom Filters (with 10 bits) and use data block hash index~\cite{rocksdb-tuning, rocksdb-point}, using the \texttt{OptimizeForPointLookup()} option.
SplinterDB employs quotient filters, which we enable.

We configure \faster with fixed $1$GiB hash index ($8$ tag bits) and $1.75$GiB \hlog in-memory region ($90\%$ mutable like in~\cite{FASTER}).
For most experiments, we replace \faster's original scan-based compaction with \sys's lookup-based one, to avoid exceeding the memory budget during compaction process.

\bsubsection{\sys Configuration.}
Unless otherwise noted, we configure \sys as per our guidelines (see Section~\ref{sec:f2-implementation-usage}).
The mutable region of the hot log is set to $90\%$ of the log in-memory region.
We use $512$MiB memory budget ($4$M hash buckets) for the hot-log index and assign $512$MiB to read-cache. We only use $64$MiB for the in-memory region of the cold-log and configure the cold-log index to use $256$B hash chunks that are indexed in-memory using another $64$MiB.
The remaining budget ($\approx1.75$GiB) is mostly used for the in-memory region of the hot log.
To trigger both hot-cold and cold-cold compactions, we set the disk budget of the hot (cold) record log to $125\%$ of database size: $5$GiB ($35$GiB) for YCSB and $7.5$GiB ($42.5$GiB) for MixGraph.
Note that we use the same \sys configuration for all workloads to perform a \emph{fair} comparison against the baselines.

\subsection{Comparing \sys with the Original \faster}
\label{sec:eval-f2-impact}
 
\fbsubsection{Lookup- vs Scan-based Compaction.}
We compare \sys's lookup-based compaction with our original scan-based compaction, when compacting $2$GiB worth of records from a single log (to its tail).
As shown in Figure~\ref{fig:lookup-vs-scan-compaction}, lookup-based compaction can lead to faster compaction times, i.e., $1.8\times$ when using a single thread, or $5.2\times$ when using the same disk bandwidth (with $4$ threads).
More importantly, compactions consume $25\times$ less memory: we measure the peak memory utilization during lookup-based compaction at $120$MiB, compared to $3$GiB for scan-based compaction.
This is because the lookup-based approach only stores a fixed amount of data in-memory (i.e., several log pages), and not a growing temporary memory buffer of live records (as the scan-based approach does).

\bsubsection{Impact of Read-Cache.}
We now evaluate the impact of \sys's read-cache in system throughput.
We observe that compared to the default $512$MiB read-cache, \sys can deliver up to $19\%$ ($27\%$) higher throughput for read-heavy YCSB-B (YCSB-C), if read-cache is configured properly.
Moreover, for the read-only YCSB-C, we see that even a small cache (of $256$MiB) can provide almost a $8.3\times$ speedup over not using one.
Interestingly, for read-cache sizes of $1.5$GiB (or more), an (increasingly larger) number of cold-resident read-hot records can now be stored in-memory.
This occurs because the read-cache is now able to keep in-memory an (increasingly larger) subset of read-hot records that actually reside in the cold log.

\vspace{-1ex}
\subsection{Comparison to Baselines}
\label{sec:eval-baselines}

\begin{figure}[t]
    \centering
    \includegraphics[width=1\linewidth,trim={0 1mm 0 0},clip]{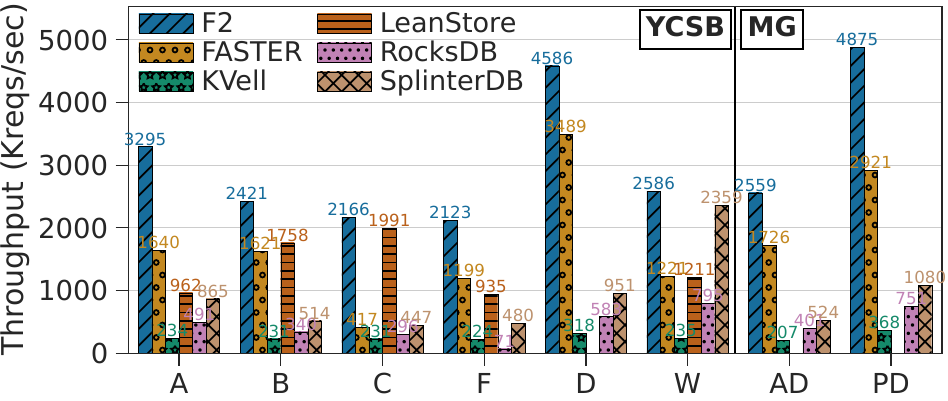}
    \caption{Throughput of \sys and baselines on all workloads.}
    \label{fig:e2e-zipfian}
    \label{fig:e2e}
    \vspace{-2.5ex}
\end{figure}

\begin{figure}[t]
    \centering
    \includegraphics[width=0.95\linewidth,trim={0 7.45cm 0 0.1cm},clip]{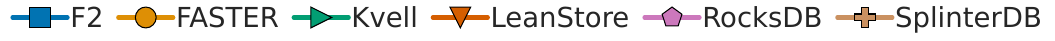}
    \begin{minipage}{0.48\linewidth}
        \centering
        \includegraphics[width=\textwidth]{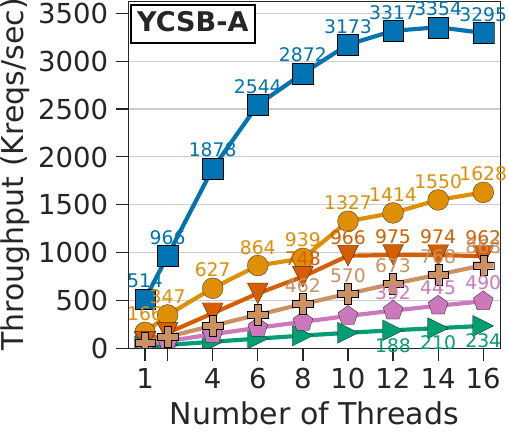}
        \label{fig:scalability-ycsbA}
    \end{minipage}
    \hspace{0.01\linewidth}
    \begin{minipage}{0.48\linewidth}
        \centering
        \includegraphics[width=\textwidth]{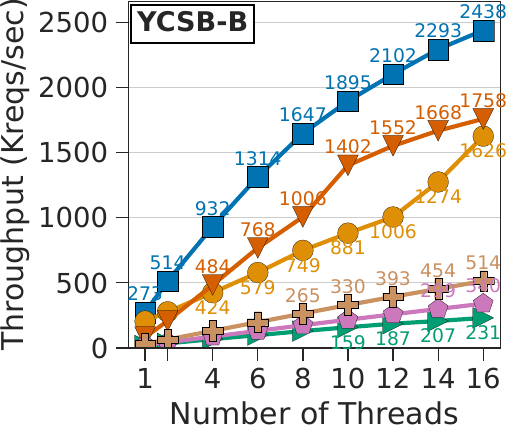}
        \label{fig:scalability-ycsbB}
    \end{minipage}
    \vspace{-3.5ex}
    \caption{Thread scaling of \sys and baselines systems for Zipfian YCSB-A (left) and YCSB-B (right).}
    \label{fig:scalability}
    \vspace{-0.5ex}
\end{figure}

\fbsubsection{System Throughput.}
We now evaluate \sys performance compared to the baseline systems, when using all $16$ CPU cores and with the available memory as $10\%$ of the workload dataset (i.e., $3$GiB).
Figure~\ref{fig:e2e} shows system throughput (Kops/sec) for YCSB and MixGraph workloads.
\sys outperforms \faster ($2.1\times$), LeanStore ($2\times$), KVell ($11.9\times$), SplinterDB ($4.6\times$), and  RocksDB ($11.8\times$) on average.
For update-heavy \mbox{YCSB-A}, \sys speedups stem from its tiered log-structured design, which enables fast ingestion speeds, while keeping the most-frequently accessed keys in the in-memory part of hot-log.
For read-heavy MixGraph (YCSB-B, YCSB-C), \sys is $1.5$--$4.8\times$ faster than \faster and SplinterDB, while matching LeanStore for YCSB (its driver does not support MixGraph), mostly due to its read-cache and smaller hash chains (i.e., less read I/O).

KVell's poor performance is attributed to its large in-memory index, parts of which are continuously being swapped out to disk.
LeanStore's page-oriented buffer pool cannot keep the hot records in-memory, as these are not always clustered together in the same page; however, it saturates the I/O bandwidth, achieving good performance.
\faster's "death-spiral" effect is observed on update-intensive YCSB-A/F/W workloads: i.e., continuous compaction of write-cold records to log tail, leading to the eviction of hot ones.

Overall, \sys provides meaningful speedups ($2$--$11.9\times$) across many YCSB and real-world MixGraph workloads.

\bsubsection{Latency.}
We now evaluate \sys request latency compared to baselines. We use a single thread to avoid any potential additional disk controller delays caused by larger I/O queue depth.
Table~\ref{tab:e2e-latency} lists the average latency (in microseconds) for three YCSB workloads.
We observe that \sys's average read/write request latency is on par with the best baselines, FASTER and SplinterDB, while achieving $1.4\times$ ($3.5\times$) lower read (write) latency compared to RocksDB.

\begin{table}[t]
\centering
\footnotesize
\addtolength{\tabcolsep}{-0.4em}
\caption{Average user request latency (us) for \sys and baselines.}
\vspace{-.5ex}
\resizebox{\linewidth}{!}{
\begin{tabular}{l|l|cc|cc|c}
\hline
\multicolumn{2}{c|}{\textbf{Workload}} & \textbf{FASTER} & \textbf{LeanStore} & \textbf{RocksDB} & \textbf{SplinterDB} & \textbf{F2} \\
\hline
\multirow{3}{*}{\rotatebox{90}{\textbf{Read}}} & YCSB-A & 41.5 & 254.3 & 53.0 & 42.1 & 41.7 \\
& YCSB-B & 47.2 & 122.0 & 50.7 & 42.1 & 45.3 \\
& YCSB-W & 40.9 & 311.8 & 63.1 & 58.8 & 39.2 \\
\hline
\hline
\multirow{3}{*}{\rotatebox{90}{\textbf{Write}}} & YCSB-A & 4.2 & 262.8 & 10.3 & 3.9 & 3.4 \\
& YCSB-B & 4.8 & 128.8 & 6.7 & 3.9 & 3.9  \\
& YCSB-W & 4.1 & 324.9 & 17.3 & 3.5 & 3.3  \\
\hline
\end{tabular}
}
\label{tab:e2e-ioamp}
\vspace{-3ex}
\end{table}

\begin{table}[t]
\centering
\footnotesize
\addtolength{\tabcolsep}{-0.4em}
\caption{Disk read (RA) and write (WA) amplification for \sys and baselines on YCSB-A, YCSB-B, YCSB-W, and MG-PD.}
\vspace{-.5ex}
\resizebox{\linewidth}{!}{
\begin{tabular}{l|l|ccc|cc|c}
\hline
\multicolumn{2}{c|}{\textbf{Workload}} & \textbf{FASTER} & \textbf{KVell} & \textbf{LeanStore} & \textbf{RocksDB} & \textbf{SplinterDB} & \textbf{F2} \\
\hline
\multirow{3}{*}{\rotatebox{90}{\textbf{RA}}} & YCSB-A & 7.23 & 91.93 & 66.87 & 21.47 & 17.09 & 6.41 \\
& YCSB-B & 5.03 & 48.71 & 35.22 & 16.51 & 15.75 & 5.5 \\
& YCSB-W & 38.6 & 95.32 & 72.23 & 109.12 & 52.97 & 53.68 \\
& MG-PD & 1.91 & 33.72 & - & 5.23 & 4.61 & 1.84 \\
\hline
\hline
\multirow{3}{*}{\rotatebox{90}{\textbf{WA}}} & YCSB-A & 2.62 & 31.17 & 34.47 & 5.28 & 2.18 & 1.23 \\
& YCSB-B & 1.21 & 33.79 & 38.73 & 2.64 & 2.52 & 1.77 \\
& YCSB-W & 1.75 & 31.48 & 32.87 & 5.81 & 1.85 & 1.48 \\
& MG-PD & 2.26 & 32.01 & - & 2.52 & 2.38 & 2.43 \\
\hline
\end{tabular}
}
\label{tab:e2e-latency}
\end{table}

\bsubsection{I/O Amplification.}
Table~\ref{tab:e2e-ioamp} lists disk read and write amplification for several as measured by \texttt{proc/io}.
We observe that \sys reads $2.5$--$2.9\times$ less bytes from disk compared to SplinterDB for read-intensive YCSB-B and MG-PD, due to the in-memory region of \sys's hot log and its read-cache that provide immediate access to hot records.
For update-intensive YCSB-A/W, we see that \sys writes $1.3$-$1.7\times$ fewer bytes to disk, compared to the best-performing system, SplinterDB.
This is attributed to the in-place update region of \sys's hot log, which avoids writing stale values to disk for write-hot keys, as well as its log-structured design, which aggregates multiple records (or hash chunks), before writing them to disk in larger ($4$KiB) blocks.
Note that even with \sys's cold-log index writing hash chunks to disk, \sys writes a comparable number of bytes to disk compared to \faster, and is more disk-friendly compared to LSM-based systems.
Unsurprisingly, page-oriented designs, i.e., LeanStore, KVell, incur high write amplification (i.e., $\ge30\times$ for 8B key, 100B value records).

Overall, \sys achieves minimal disk wear, i.e., $1.3$--$3.9\times$ lower write amplification than LSM-based stores, on average.

\bsubsection{Thread Scaling.}
We now evaluate system throughput by varying the number of threads.
Figure~\ref{fig:scalability} shows throughput for YCSB-A and YCSB-B.
For YCSB-A, we observe that \sys scales linearly from $1$ to $6$ threads, but between $10$-$12$ threads the scaling flattens out.
LeanStore manages to saturate the disk bandwidth with $10$ threads; adding more threads do not result in better performance due to inefficient record caching.
Both RocksDB and SplinterDB show good scaling, with the latter showing superior ingestion behavior, yet they cannot saturate disk bandwidth.
Finally, for YCSB-B, \sys, LeanStore and \faster achieve good thread scaling, and saturate $85$--$90\%$ of the disk bandwidth with $16$ threads.

\subsection{Varying the Workload Skewness}
\label{sec:eval-skewness}

\begin{figure}[t]
    \centering
    \begin{subfigure}[b]{0.46\linewidth}
         \centering
         \includegraphics[width=\textwidth]{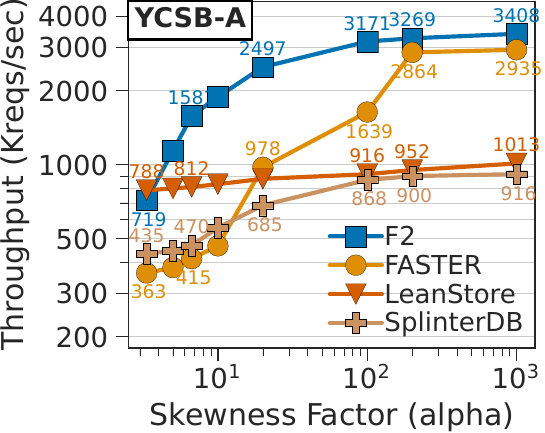}
    \end{subfigure}
    \hspace{0.01\linewidth}
    \begin{subfigure}[b]{0.46\linewidth}
         \centering
         \includegraphics[width=\textwidth]{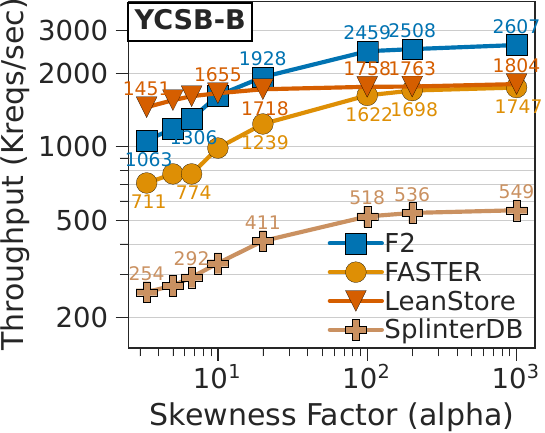}
    \end{subfigure}
    \caption{Throughput on YCSB-A (left), YCSB-B (right), when varying Zipf Skewness Factor ($\alpha$). Axes in log scale.}
    \label{fig:hs-sensitivity}
\end{figure}

\sys targets larger-than-memory workloads with skewed key distribution.
To better understand how \sys behaves under different skewed distributions, we experiment with varying the Zipfian skewness factor $\alpha$, from $3$ to $1000$ (higher values mean more skewed accesses).
When $\alpha$=$100$ (YCSB default), $90\%$ of accesses go to $18\%$ of records; for $\alpha$=$10$,~$90\%$ of accesses go to $33\%$ of records.
As before, we use $16$ threads, and set the memory budget to $3$GiB.
Figure~\ref{fig:hs-sensitivity} shows total system throughput for YCSB-A, YCSB-B (axes in log scale).

For high skewness factors ($\alpha \ge 200$), \sys performs $3.4\times$
($1.4\times$) better for YCSB-A (YCSB-B) compared to LeanStore, and matches or outperforms \faster (e.g., by $1.5\times$ for YCSB-B), due to its effective physical separation of hot and cold records, minimizing compaction and user-related disk operations.
As we decrease workload skewness ($\alpha \le 50$), \sys's performance gracefully degrades, as the hot set now spills over to disk (and subsequently to cold-log for $\alpha \le 20$).
For update-intensive YCSB-A, \sys's fast ingestion capability and efficient lookup-based compaction manage to retain superior performance, even for less-skewed workload (i.e., $\alpha=3$).
For read-intensive YCSB-B, \sys's performance degrades, as for more requests now \sys needs to issue two additional I/O (i.e., cold-log resident records).
Yet, even with its moderately-sized read-cache (i.e., $512$MiB), \sys outperforms the original \faster by almost $2\times$.

\vspace{-1ex}
\subsection{Varying Memory Budgets}
\label{sec:eval-mem-budget}

\sys aims for high-performance even when deployed on constrained memory environments.
Here, we experiment with varying memory budgets, ranging from $750$MiB to $7.5$GiB ($2.5$--$25\%$ of our $250$M dataset), using YCSB-A and YCSB-B.
We use $16$ threads, and configure each system to adhere with the memory limit (we also impose this limit via Linux \texttt{cgroups}).
Specifically, for \sys we only change the size of the in-memory region of the hot log based on the available budget, while keeping everything else constant (e.g., read-cache).
When operating on the lowest $750$MiB budget, we disable the read-cache (to make space for the hot-log index and in-memory hot-log region).
For LeanStore/SplinterDB (\faster) we adjust the size of the in-memory buffer-pool/cache (log) based on memory budget.

\begin{figure}[t]
    \centering
    \begin{subfigure}[b]{0.46\linewidth}
         \centering
         \includegraphics[width=\textwidth]{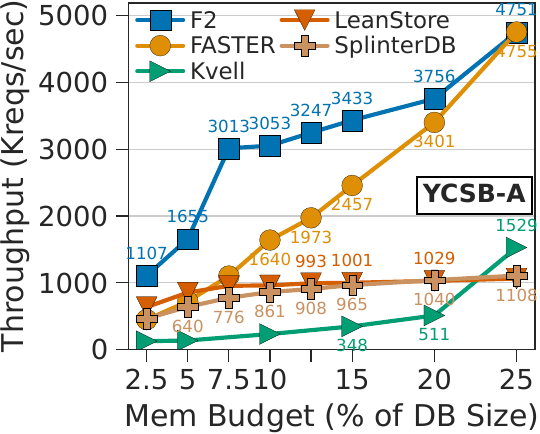}
         \label{fig:ds-ycsbA}
    \end{subfigure}
    \hspace{0.02\linewidth}
    \begin{subfigure}[b]{0.46\linewidth}
         \centering
         \includegraphics[width=\textwidth]{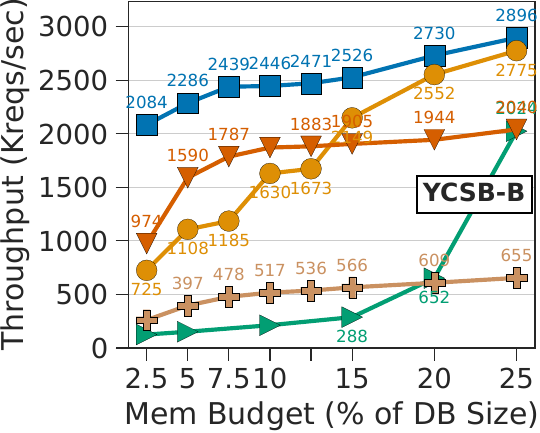}
         \label{fig:ds-ycsbB}
    \end{subfigure}
    \vspace{-3ex}
    \caption{Throughput on memory-scarce environments. Memory budget is $2.5\%-25\%$ of database size ($30$GiB). Zipfian YCSB-A (left), YCSB-B (right). X-axis in log scale.}
    \label{fig:ds-sensitivity}
    \vspace{-.5ex}
\end{figure}

Figure~\ref{fig:ds-sensitivity} shows throughput of the best systems as we increase the available memory budget.
Given a minuscule memory budget of $750$MiB ($2.5\%$ of $30$GiB), \sys achieves $36\%$ ($83\%$) of the performance for YCSB-A (YCSB-B) when given $4\times$ more memory, while still performing $1.73\times$ ($2.14\times$) better than the best system on the same budget, LeanStore.
When using such small budgets ($2.5$-$5\%$) on YCSB-A even the hottest records do not fit in-memory, forcing \sys to perform mostly I/O operations.
Once we give slightly larger budget of $2.25$GiB ($\ge7.5\%$), \sys sees a performance jump of $2\times$, as most hot records are now in-memory, leading to $3.1$-$4.7\times$ better throughput compared to LeanStore/SplinterDB, while always matching or outperforming \faster.
On YCSB-B, \sys quickly saturates the disk bandwidth at $2.25$GiB ($7.5\%$), and any further gains stem from serving in-memory hot records (i.e., due to larger in-memory hot-log size).
LeanStore and \faster, while slow on small budgets, manage to perform well on higher ones, with \faster even matching \sys's performance for budgets of $\ge6$GiB ($20$-$25\%$).

In summary, \sys outperforms the best systems on small-moderate memory budgets and matches their performance on larger ones.

\balance

\vspace{-1ex}
\section{Related Work}

\fbsubsection{Memory-efficient Designs.}
Log-structured Merge (LSM) Trees~\cite{lsm} designs prioritize memory-efficiency, and they can store TBs of data.
They have been widely adopted as the storage layer for many popular key-value stores~\cite{leveldb, rocksdb, pebblesdb, splinterdb}.
Researchers have proposed many optimizations for LSM-Trees, including better compaction algorithms~\cite{spooky, pebblesdb, lsm-compaction} (or policies for different tiers~\cite{dostoevsky, monkey}), and smaller and more-performant filters~\cite{chucky, rosetta-filter, splinterdb-mapplets}.
State-of-the-art LSM-based systems, like SplinterDB~\cite{splinterdb, splinterdb-mapplets}, integrate additional optimizations aimed at improving concurrency and I/O bandwidth utilization, like STB$\epsilon$-tree, flush-then-compact compaction, quotient mapplets~\cite{quotient, splinterdb-mapplets}, which further reduce write amplification.

\noindent
\textbf{B-Tree based Designs.}
KVell~\cite{kvell} uses a B-Tree to map every key to a page offset on disk.
Disk pages are cached in-memory using a dedicated page cache.
In KVell, each thread is responsible for handling requests only for a subset of the key space, eliminating thread contention.
LeanStore~\cite{leanstore} is optimized for modern NVMe SSDs and multi-core CPUs, and uses a B-Tree alongside an in-memory page buffer manager to support larger-than-memory workloads.
Its key idea is pointer swizzling: cached pages are directly accessible via pointers, avoiding the indirection necessary in traditional buffer manager designs.
It employs additional techniques (e.g., optimistic locking, contention split) to improve concurrency~\cite{leanstore_evolution}.
Bw-Tree~\cite{bw-tree} uses log-structured writes using delta records for pages, as discussed in Section~\ref{sec:background}.
Recent work~\cite{2-tree} proposes a migration process that clusters hot (cold) records together to create hot (cold) pages, by moving records across pages, improving caching effectiveness.

\vspace{-1ex}
\section{Conclusion}

This paper describes our journey from the original \faster library to \sys (for FASTER v2), an evolved key-value store design that targets large skewed workloads. \sys addresses the limitations of existing systems that prevent them from serving such workloads effectively.
\sys is open-sourced and available as part of the FASTER project.

\bibliographystyle{plain}
\bibliography{papers,urls}

\begin{thebibliography}{10}

\bibitem{leanstore_evolution}
Adnan Alhomssi, Michael Haubenschild, and Viktor Leis.
\newblock The evolution of leanstore.
\newblock In {\em BTW 2023}, pages 259--281. Gesellschaft f{\"u}r Informatik eV, 2023.

\bibitem{10.14778/1687553.1687590}
M.~H. Ali, C.~Gerea, B.~S. Raman, B.~Sezgin, T.~Tarnavski, T.~Verona, P.~Wang, P.~Zabback, A.~Ananthanarayan, A.~Kirilov, M.~Lu, A.~Raizman, R.~Krishnan, R.~Schindlauer, T.~Grabs, S.~Bjeletich, B.~Chandramouli, J.~Goldstein, S.~Bhat, Ying Li, V.~Di~Nicola, X.~Wang, David Maier, S.~Grell, O.~Nano, and I.~Santos.
\newblock Microsoft cep server and online behavioral targeting.
\newblock {\em Proc. VLDB Endow.}, 2(2):1558–1561, August 2009.

\bibitem{skew}
Berk Atikoglu, Yuehai Xu, Eitan Frachtenberg, Song Jiang, and Mike Paleczny.
\newblock Workload analysis of a large-scale key-value store.
\newblock In {\em Proceedings of the 12th ACM SIGMETRICS/PERFORMANCE Joint International Conference on Measurement and Modeling of Computer Systems}, SIGMETRICS '12, page 53–64, New York, NY, USA, 2012. Association for Computing Machinery.

\bibitem{linearizability}
Hagit Attiya and Jennifer~L. Welch.
\newblock Sequential consistency versus linearizability.
\newblock {\em ACM Trans. Comput. Syst.}, 12(2):91–122, May 1994.

\bibitem{azure-stream-analytics}
{Azure Stream Analytics}.
\newblock \url{https://azure.microsoft.com/en-us/products/stream-analytics}, February 2025.

\bibitem{burckhardt2022netherite}
Sebastian Burckhardt, Badrish Chandramouli, Chris Gillum, David Justo, Konstantinos Kallas, Connor McMahon, Christopher~S Meiklejohn, and Xiangfeng Zhu.
\newblock Netherite: Efficient execution of serverless workflows.
\newblock {\em Proceedings of the VLDB Endowment}, 15(8):1591--1604, 2022.

\bibitem{burckhardt2025netherite}
Sebastian Burckhardt, Badrish Chandramouli, Chris Gillum, David Justo, Konstantinos Kallas, Connor McMahon, Christopher~S Meiklejohn, and Xiangfeng Zhu.
\newblock Netherite: efficient execution of serverless workflows.
\newblock {\em The VLDB Journal}, 34(2):25, 2025.

\bibitem{adf}
Sebastian Burckhardt, Chris Gillum, David Justo, Konstantinos Kallas, Connor McMahon, and Christopher~S Meiklejohn.
\newblock Durable functions: Semantics for stateful serverless.
\newblock {\em Proceedings of the ACM on Programming Languages}, 5(OOPSLA):1--27, 2021.

\bibitem{rocksdb}
Zhichao Cao and Siying Dong.
\newblock Characterizing, modeling, and benchmarking rocksdb key-value workloads at facebook.
\newblock In {\em 18th USENIX Conference on File and Storage Technologies (FAST’20)}, 2020.

\bibitem{10.1109/ICDE.2012.55}
Badrish Chandramouli, Jonathan Goldstein, and Songyun Duan.
\newblock Temporal analytics on big data for web advertising.
\newblock In {\em Proceedings of the 2012 IEEE 28th International Conference on Data Engineering}, ICDE '12, page 90–101, USA, 2012. IEEE Computer Society.

\bibitem{FASTER}
Badrish Chandramouli, Guna Prasaad, Donald Kossmann, Justin Levandoski, James Hunter, and Mike Barnett.
\newblock {FASTER}: A concurrent key-value store with in-place updates.
\newblock In {\em Proceedings of the 2018 International Conference on Management of Data}, pages 275--290, 2018.

\bibitem{faster-demo}
Badrish Chandramouli, Guna Prasaad, Donald Kossmann, Justin Levandoski, James Hunter, and Mike Barnett.
\newblock {FASTER}: an embedded concurrent key-value store for state management.
\newblock {\em Proceedings of the VLDB Endowment}, 11(12):1930--1933, 2018.

\bibitem{chang2008bigtable}
Fay Chang, Jeffrey Dean, Sanjay Ghemawat, Wilson~C Hsieh, Deborah~A Wallach, Mike Burrows, Tushar Chandra, Andrew Fikes, and Robert~E Gruber.
\newblock Bigtable: A distributed storage system for structured data.
\newblock {\em ACM Transactions on Computer Systems (TOCS)}, 26(2):1--26, 2008.

\bibitem{cgroup}
{Control Group v2}.
\newblock \url{https://docs.kernel.org/admin-guide/cgroup-v2.html}, October 2023.

\bibitem{splinterdb-mapplets}
Alex Conway, Mart\'{\i}n Farach-Colton, and Rob Johnson.
\newblock Splinterdb and maplets: Improving the tradeoffs in key-value store compaction policy.
\newblock {\em Proc. ACM Manag. Data}, 1(1), may 2023.

\bibitem{splinterdb}
Alexander Conway, Abhishek Gupta, Vijay Chidambaram, Martin Farach-Colton, Richard Spillane, Amy Tai, and Rob Johnson.
\newblock {SplinterDB}: Closing the bandwidth gap for {NVMe} {Key-Value} stores.
\newblock In {\em 2020 USENIX Annual Technical Conference (USENIX ATC 20)}, pages 49--63. USENIX Association, July 2020.

\bibitem{ycsb}
Brian~F Cooper, Adam Silberstein, Erwin Tam, Raghu Ramakrishnan, and Russell Sears.
\newblock Benchmarking cloud serving systems with ycsb.
\newblock In {\em Proceedings of the 1st ACM symposium on Cloud computing}, pages 143--154, 2010.

\bibitem{monkey}
Niv Dayan, Manos Athanassoulis, and Stratos Idreos.
\newblock Monkey: Optimal navigable key-value store.
\newblock In {\em Proceedings of the 2017 ACM International Conference on Management of Data}, SIGMOD '17, page 79–94, New York, NY, USA, 2017. Association for Computing Machinery.

\bibitem{dostoevsky}
Niv Dayan and Stratos Idreos.
\newblock Dostoevsky: Better space-time trade-offs for lsm-tree based key-value stores via adaptive removal of superfluous merging.
\newblock In {\em Proceedings of the 2018 International Conference on Management of Data}, SIGMOD '18, page 505–520, New York, NY, USA, 2018. Association for Computing Machinery.

\bibitem{chucky}
Niv Dayan and Moshe Twitto.
\newblock Chucky: A succinct cuckoo filter for lsm-tree.
\newblock In {\em Proceedings of the 2021 International Conference on Management of Data}, SIGMOD '21, page 365–378, New York, NY, USA, 2021. Association for Computing Machinery.

\bibitem{spooky}
Niv Dayan, Tamar Weiss, Shmuel Dashevsky, Michael Pan, Edward Bortnikov, and Moshe Twitto.
\newblock Spooky: granulating lsm-tree compactions correctly.
\newblock {\em Proc. VLDB Endow.}, 15(11):3071–3084, July 2022.

\bibitem{skimpystash}
Biplob Debnath, Sudipta Sengupta, and Jin Li.
\newblock Skimpystash: Ram space skimpy key-value store on flash-based storage.
\newblock In {\em Proceedings of the 2011 ACM SIGMOD International Conference on Management of Data}, SIGMOD '11, page 25–36, New York, NY, USA, 2011. Association for Computing Machinery.

\bibitem{anti-caching}
Justin DeBrabant, Andrew Pavlo, Stephen Tu, Michael Stonebraker, and Stan Zdonik.
\newblock Anti-caching: A new approach to database management system architecture.
\newblock {\em Proceedings of the VLDB Endowment}, 6(14):1942--1953, 2013.

\bibitem{rocksdb-evolution}
Siying Dong, Andrew Kryczka, Yanqin Jin, and Michael Stumm.
\newblock Rocksdb: Evolution of development priorities in a key-value store serving large-scale applications.
\newblock {\em ACM Trans. Storage}, 17(4), October 2021.

\bibitem{Cloudlab}
Dmitry Duplyakin, Robert Ricci, Aleksander Maricq, Gary Wong, Jonathon Duerig, Eric Eide, Leigh Stoller, Mike Hibler, David Johnson, Kirk Webb, Aditya Akella, Kuangching Wang, Glenn Ricart, Larry Landweber, Chip Elliott, Michael Zink, Emmanuel Cecchet, Snigdhaswin Kar, and Prabodh Mishra.
\newblock The design and operation of cloudlab.
\newblock In {\em Proceedings of the 2019 USENIX Conference on Usenix Annual Technical Conference}, USENIX ATC '19, page 1–14, USA, 2019. USENIX Association.

\bibitem{faster-github}
{FASTER: Fast persistent recoverable log and key-value store + cache}.
\newblock \url{https://github.com/microsoft/FASTER}, October 2023.

\bibitem{leanstore-nvme}
Gabriel Haas and Viktor Leis.
\newblock What modern nvme storage can do, and how to exploit it: High-performance i/o for high-performance storage engines.
\newblock {\em Proc. VLDB Endow.}, 16(9):2090–2102, May 2023.

\bibitem{huynh2021endure}
Andy Huynh, Harshal~A Chaudhari, Evimaria Terzi, and Manos Athanassoulis.
\newblock Endure: a robust tuning paradigm for lsm trees under workload uncertainty.
\newblock {\em arXiv preprint arXiv:2110.13801}, 2021.

\bibitem{rocksdb-point}
{Improving Point-Lookup Using Data Block Hash Index}.
\newblock \url{https://rocksdb.org/blog/2018/08/23/data-block-hash-index.html}, August 2018.

\bibitem{leanstore}
Viktor Leis, Michael Haubenschild, Alfons Kemper, and Thomas Neumann.
\newblock Leanstore: In-memory data management beyond main memory.
\newblock In {\em 2018 IEEE 34th International Conference on Data Engineering (ICDE)}, pages 185--196. IEEE, 2018.

\bibitem{kvell}
Baptiste Lepers, Oana Balmau, Karan Gupta, and Willy Zwaenepoel.
\newblock Kvell: the design and implementation of a fast persistent key-value store.
\newblock In {\em Proceedings of the 27th ACM Symposium on Operating Systems Principles}, pages 447--461, 2019.

\bibitem{bw-tree}
Justin Levandoski, David Lomet, and Sudipta Sengupta.
\newblock The bw-tree: A b-tree for new hardware platforms.
\newblock In {\em 2013 IEEE 29th International Conference on Data Engineering (ICDE)}. IEEE, April 2013.

\bibitem{hotcold-data}
Justin~J. Levandoski, Per-Åke Larson, and Radu Stoica.
\newblock Identifying hot and cold data in main-memory databases.
\newblock In {\em 2013 IEEE 29th International Conference on Data Engineering (ICDE)}, pages 26--37, 2013.

\bibitem{leveldb}
{LevelDB}.
\newblock \url{https://github.com/google/leveldb}, October 2023.

\bibitem{silt}
Hyeontaek Lim, Bin Fan, David~G. Andersen, and Michael Kaminsky.
\newblock Silt: A memory-efficient, high-performance key-value store.
\newblock In {\em Proceedings of the Twenty-Third ACM Symposium on Operating Systems Principles}, SOSP '11, page 1–13, New York, NY, USA, 2011. Association for Computing Machinery.

\bibitem{lsm-survey}
Chen Luo and Michael~J. Carey.
\newblock Lsm-based storage techniques: A survey.
\newblock {\em The VLDB Journal}, 29(1):393–418, jul 2019.

\bibitem{rosetta-filter}
Siqiang Luo, Subarna Chatterjee, Rafael Ketsetsidis, Niv Dayan, Wilson Qin, and Stratos Idreos.
\newblock Rosetta: A robust space-time optimized range filter for key-value stores.
\newblock In {\em Proceedings of the 2020 ACM SIGMOD International Conference on Management of Data}, SIGMOD '20, page 2071–2086, New York, NY, USA, 2020. Association for Computing Machinery.

\bibitem{larger-than-memory}
Lin Ma, Joy Arulraj, Sam Zhao, Andrew Pavlo, Subramanya~R. Dulloor, Michael~J. Giardino, Jeff Parkhurst, Jason~L. Gardner, Kshitij Doshi, and Stanley Zdonik.
\newblock Larger-than-memory data management on modern storage hardware for in-memory oltp database systems.
\newblock In {\em Proceedings of the 12th International Workshop on Data Management on New Hardware}, DaMoN '16, New York, NY, USA, 2016. Association for Computing Machinery.

\bibitem{disk-wear}
Changwoo Min, Kangnyeon Kim, Hyunjin Cho, Sang-Won Lee, and Young~Ik Eom.
\newblock Sfs: Random write considered harmful in solid state drives.
\newblock In {\em Proceedings of the 10th USENIX Conference on File and Storage Technologies}, FAST'12, page~12, USA, 2012. USENIX Association.

\bibitem{nuget}
{NuGet Gallery}.
\newblock \url{https://nuget.org/}, July 2025.

\bibitem{lsm}
Patrick O’Neil, Edward Cheng, Dieter Gawlick, and Elizabeth O’Neil.
\newblock The log-structured merge-tree (lsm-tree).
\newblock {\em Acta Informatica}, 33:351--385, 1996.

\bibitem{quotient}
Prashant Pandey, Michael~A. Bender, Rob Johnson, and Rob Patro.
\newblock A general-purpose counting filter: Making every bit count.
\newblock In {\em Proceedings of the 2017 ACM International Conference on Management of Data}, SIGMOD '17, page 775–787, New York, NY, USA, 2017. Association for Computing Machinery.

\bibitem{pebblesdb}
Pandian Raju, Rohan Kadekodi, Vijay Chidambaram, and Ittai Abraham.
\newblock Pebblesdb: Building key-value stores using fragmented log-structured merge trees.
\newblock In {\em Proceedings of the 26th Symposium on Operating Systems Principles}, pages 497--514, 2017.

\bibitem{rocksdb-tuning}
{RocksDB Tuning Guide}.
\newblock \url{https://github.com/facebook/rocksdb/wiki/RocksDB-Tuning-Guide}, october 2023.

\bibitem{lsm-compaction}
Subhadeep Sarkar, Kaijie Chen, Zichen Zhu, and Manos Athanassoulis.
\newblock Compactionary: A dictionary for lsm compactions.
\newblock In {\em Proceedings of the 2022 International Conference on Management of Data}, SIGMOD '22, page 2429–2432, New York, NY, USA, 2022. Association for Computing Machinery.

\bibitem{tanenbaum-os}
Andrew~S Tanenbaum and Albert~S Woodhull.
\newblock {\em Operating systems: design and implementation}, volume~2.
\newblock Prentice Hall Englewood Cliffs, 1997.

\bibitem{silo}
Stephen Tu, Wenting Zheng, Eddie Kohler, Barbara Liskov, and Samuel Madden.
\newblock Speedy transactions in multicore in-memory databases.
\newblock In {\em Proceedings of the Twenty-Fourth ACM Symposium on Operating Systems Principles}, SOSP '13, page 18–32, New York, NY, USA, 2013. Association for Computing Machinery.

\bibitem{twitter-analysis}
Juncheng Yang, Yao Yue, and K.~V. Rashmi.
\newblock A large scale analysis of hundreds of in-memory cache clusters at twitter.
\newblock In {\em 14th USENIX Symposium on Operating Systems Design and Implementation (OSDI 20)}, pages 191--208. USENIX Association, November 2020.

\bibitem{treeline}
Geoffrey~X. Yu, Markos Markakis, Andreas Kipf, Per-\r{A}ke Larson, Umar~Farooq Minhas, and Tim Kraska.
\newblock Treeline: An update-in-place key-value store for modern storage.
\newblock {\em Proc. VLDB Endow.}, 16(1):99–112, sep 2022.

\bibitem{adoc}
Jinghuan Yu, Sam~H. Noh, Young-ri Choi, and Chun~Jason Xue.
\newblock Adoc: automatically harmonizing dataflow between components in log-structured key-value stores for improved performance.
\newblock In {\em Proceedings of the 21st USENIX Conference on File and Storage Technologies}, FAST'23, USA, 2023. USENIX Association.

\bibitem{zhang2021memory}
Huanchen Zhang.
\newblock Memory-efficient search trees for database management systems.
\newblock In {\em Proceedings of the 2021 International Conference on Management of Data}, SIGMOD '21, page~9, New York, NY, USA, 2021. Association for Computing Machinery.

\bibitem{2-tree}
Xinjing Zhou, Xiangyao Yu, Goetz Graefe, and Michael Stonebraker.
\newblock Two is better than one: The case for 2-tree for skewed data sets.
\newblock In {\em 13th Conference on Innovative Data Systems Research, {CIDR} 2023, Amsterdam, Online Proceedings}, 2023.

\end{thebibliography}

\end{document}